\documentclass[useAMS,usegraphicx]{mn2e}
\usepackage{aas_macros}
\usepackage{times}
\usepackage{longtable}
\usepackage[small,bf,singlelinecheck=off]{caption}

\def\gs{\mathrel{\raise0.35ex\hbox{$\scriptstyle >$}\kern-0.6em
\lower0.40ex\hbox{{$\scriptstyle \sim$}}}}
\def\ls{\mathrel{\raise0.35ex\hbox{$\scriptstyle <$}\kern-0.6em
\lower0.40ex\hbox{{$\scriptstyle \sim$}}}}

\title[Multi-wavelength demographics of 450$\,\mu$m-sources]
{\vspace{-0.5cm}The SCUBA-2 Cosmology Legacy Survey: demographics of the 450$\,\mu$m-population}
\vspace{-0.5em}
\author[I.G.~Roseboom et al.]
{\parbox{\textwidth}{\raggedright I.G.~Roseboom,$^{1}$\thanks{E-mail: \texttt{igr@roe.ac.uk}}
J.S.~Dunlop,$^{1}$
M.~Cirasuolo,$^{1}$
J.E.~Geach,$^{2}$
I.~Smail,$^{3}$
M.~Halpern,$^{4}$
P.~van~der~Werf,$^{5}$
O.~Almaini,$^{6}$
V.~Arumugam,$^{7}$
V.~Asboth,$^{4}$
R.~Auld,$^{8}$
A.~Blain,$^{9}$
M.N~Bremer,$^{10}$
J.~Bock,$^{11,12}$
R.~Bowler,$^{1}$
F.~Buitrago,$^{1}$ 
E.~Chapin,$^{13,4}$
S.~Chapman,$^{14}$
A.~Chrysostomou$^{15,16}$
C.~Clarke,$^{17}$
A.~Conley,$^{18}$
K.~E.~K.~Coppin,$^{2}$
A.L.R~Danielson,$^{3}$
D.~Farrah,$^{19}$
J.~Glenn,$^{18}$
E.~Hatziminaoglou,$^{20}$
E.~Ibar,$^{7}$
R.J.~Ivison,$^{7,1}$
T.~Jenness$^{21, 16}$
E.~van Kampen$^{20}$
A.~Karim,$^{3}$
T.~Mackenzie,$^{4}$
G.~Marsden,$^{4}$
R.~Meijerink,$^{24}$
M.J.~Micha{\l}owski,$^{23,1}$\thanks{FWO Pegasus Marie Curie Fellow}
S.J.~Oliver,$^{17}$
M.J~Page,$^{22}$
E.~Pearson,$^{8}$
Douglas~Scott,$^{4}$
J.M~Simpson,$^{3}$
D.J.B.~Smith,$^{15}$
M.~Spaans,$^{24}$
A.~M.~Swinbank$^{3}$
M.~Symeonidis,$^{22}$
T.~Targett,$^{1}$
E.~Valiante,$^{8}$
M.~Viero,$^{11}$
L.~Wang,$^{3}$
C.J.~Willott,$^{25}$ and
M.~Zemcov$^{11,12}$}\vspace{0.4cm}\\
\parbox{\textwidth}{\raggedright $^{1}$Institute for Astronomy, University of Edinburgh, Royal Observatory, Blackford Hill, Edinburgh EH9 3HJ, UK\\
$^{2}$Department of Physics, Ernest Rutherford Building, 3600 rue University, McGill University, Montr\'{e}al, QC, H3A 2T8, Canada\\
$^{3}$Institute for Computational Cosmology, Department of Physics, Durham University, South Road, Durham, DH1 3LE\\
$^{4}$Department of Physics \& Astronomy, University of British Columbia, 6224 Agricultural Road, Vancouver, BC V6T~1Z1, Canada\\
$^{5}$Leiden Observatory, Leiden University, P.O. box 9513, 2300 RA Leiden, The Netherlands\\
$^{6}$School of Physics and Astronomy, University of Nottingham, University Park, Nottingham, NG9 2RD\\
$^{7}$UK Astronomy Technology Centre, Royal Observatory, Blackford Hill, Edinburgh EH9 3HJ, UK\\
$^{8}$Cardiff School of Physics and Astronomy, Cardiff University, Queens Buildings, The Parade, Cardiff CF24 3AA, UK\\
$^{9}$Department of Physics \& Astronomy, University of Leicester, University Road, Leicester, LE1 7RH\\
$^{10}$H.H.~Wills Physics Laboratory, University of Bristol, Tyndall Avenue, Bristol, BS8 1TL, UK\\
$^{11}$California Institute of Technology, 1200 E. California Blvd., Pasadena, CA 91125, USA\\
$^{12}$Jet Propulsion Laboratory, 4800 Oak Grove Drive, Pasadena, CA 91109, USA\\
$^{13}$XMM SOC, ESAC, Apartado 78, 28691 Villanueva de la Canada, Madrid, Spain\\
$^{14}$Department of Physics and Atmospheric Science, Dalhousie University Halifax, NS, B3H 3J5, Canada\\
$^{15}$Centre for Astrophysics Research, University of Hertfordshire, College Lane, Hatfield, Hertfordshire AL10 9AB, UK\\
$^{16}$Joint Astronomy Centre 660 N. A{\'o}hoku Place University Park Hilo, Hawaii 96720, USA\\
$^{17}$Astronomy Centre, Dept. of Physics \& Astronomy, University of Sussex, Brighton BN1 9QH, UK\\
$^{18}$Dept. of Astrophysical and Planetary Sciences, CASA 389-UCB, University of Colorado, Boulder, CO 80309, USA\\
$^{19}$Virginia Polytechnic Institute \& State University Department of Physics, 910 Drillfield Drive, Blacksburg, VA 24061, USA\\
$^{20}$ESO, Karl-Schwarzschild-Str. 2, 85748 Garching bei M\"unchen, Germany\\
$^{21}$Astronomy Department, Cornell University, Ithaca, NY 14853, USA\\
$^{22}$Mullard Space Science Laboratory, University College London, Holmbury St Mary Dorking, Surrey RH5 6NT\\
$^{23}$Sterrenkundig Observatorium, Universiteit Gent, Krijgslaan 281 S9, 9000 Gent, Belgium\\
$^{24}$Kapteyn Astronomical Institute, University of Groningen, P.O. Box 800, NL-9700 AV Groningen, The Netherlands\\
$^{25}$Herzberg Institute of Astrophysics, National Research Council, 5071 West Saanich Rd, Victoria, BC V9E 2E7, Canada\\
\mbox{}\vspace{-1cm}
}}
\begin{document}
\date{\today}
\pagerange{\pageref{firstpage}--\pageref{lastpage}} \pubyear{2013}

\maketitle
\mbox{}\vspace{-0.8em}
\label{firstpage}
\begin{abstract}
We investigate the multi-wavelength properties of a sample of 450-$\mu$m selected sources from the SCUBA-2 Cosmology Legacy Survey (S2CLS). A total of 69 sources were identified above 4$\sigma$ in deep SCUBA-2 450-$\mu$m observations overlapping the UDS and COSMOS fields and covering 210 arcmin$^2$ to a typical depth of $\sigma_{450}=1.5$\,mJy. Reliable cross identification are found for 58 sources (84 per cent) in {\it Spitzer} and {\it Hubble Space Telescope} WFC3/IR data. The photometric redshift distribution ($dN/dz$) of 450$\,\mu$m-selected sources is presented, showing a broad peak in the redshift range $1<z<3$, and a median of $z=1.4$. Combining the SCUBA-2 photometry with {\it Herschel} SPIRE data from HerMES, the submm spectral energy distribution (SED) is examined via the use of modified blackbody fits, yielding aggregate values for the IR luminosity, dust temperature and emissivity of $\langle L_{\rm IR}\rangle=10^{12\pm0.8}\,$L$_{\odot}$, $\langle T_{\rm D}\rangle=42\pm11\,$K and $\langle \beta_{\rm D}\rangle=1.6\pm0.5$, respectively. The relationship between these SED parameters and the physical properties of galaxies is investigated, revealing correlations between $T_{\rm D}$ and $L_{\rm IR}$ and between $\beta_{\rm D}$ and both stellar mass and effective radius. The connection between star formation rate and stellar mass is explored, with 24 per cent of 450$\,\mu$m sources found to be ``star-bursts'', i.e. displaying anomalously high specific SFRs. However, both the number density and observed properties of these ``star-burst'' galaxies are found consistent with the population of normal star-forming galaxies.
\end{abstract}

\begin{keywords}
submillimetre: galaxies;  galaxies: starburst; galaxies: evolution
\end{keywords}

\section{Introduction}

A key goal for observational studies of galaxy formation has been to build a complete picture of star formation over cosmic time. Much of the effort in the last two decades has been focused on tracers in two distinct wavelength ranges: the near-to-far ultraviolet (UV) and the far-infrared/submm. These wavelengths are excellent tracers of star formation in distant galaxies; the UV continuum originates primarily from massive, short-lived stars, and the far-infrared (IR) from dust that envelops star forming regions and re-processes this UV continuum. 

Despite this close physical connection, these tracers have so far offered somewhat different views of the high redshift ($z>1$) Universe. Star-forming galaxies identified in the rest-frame UV are numerous, with moderate star-formation rates (SFRs $\sim10$\,M$_{\odot}$\,yr$^{-1}$; Madau et al.\ 1996; Reddy et al.\ 2006) while those detected in the submm (submm galaxies; SMGs) are rarer (by a factor of around $1/100$ in surface density) and have extremely high SFRs ($100$--$1000$\,M$_{\odot}$\,yr$^{-1}$; Smail, Ivison \& Blain 1997; Barger et al.\ 1998; Hughes et al.\ 1998; Chapman et al.\ 2005; Micha{\l}owski, Hjorth \& Watson 2010). Traditionally, these disparate properties have been interpreted as a natural extension of the behaviour seen in the local Universe, where UV-bright galaxies are common and typically star-forming disks, while the most IR-luminous galaxies (ULIRGs; $L_{\rm IR}>10^{12}L_{\odot}$) are rare and mostly the result of major mergers (e.g. Sanders et al. 1988; Farrah et al.\ 2001)

In recent years it has become clear that this single definition for SMGs cannot be completely accurate. While some SMGs are associated with merger events (e.g.\ Swinbank et al.\ 2009; Alaghband-Zadeh et al.\ 2012), it appears that a significant fraction of SMGs have disk-like morphologies (Targett et al.\ 2011, 2012). However, morphological studies of galaxies in the distant Universe require high resolution and sensitivity, restricting the size of samples studied in this way, and are difficult to interpret given the redshift (K-correction) and other selection effects.

Meanwhile, the role of SMGs in galaxy formation has been brought into focus by the emerging notion of a so-called ``main sequence'' for star forming galaxies. Large-scale studies of UV-selected samples have shown that, for a given redshift, the ratio of the SFR to the stellar mass (often referred to as the specific SFR) is roughly constant (Noeske et al.\ 2007; Elbaz et al.\ 2007; Daddi et al.\ 2007; Karim et al.\ 2011). This main-sequence has been taken as strong evidence of in-situ star formation (as opposed to mergers) being the dominant process in building stellar mass in galaxies at $z>1$ (Rodighiero et al. 2011), as mergers are thought to greatly enhance the specific SFR (e.g. Mihos \& Hernquist 1994). The specific star formation rate of SMGs has thus become of much interest, with variations in stellar mass estimates leading to opposing claims of SMGs lying off (Hainline et al.\ 2011) and on the main sequence (Micha{\l}owski et al.\ 2012).

Interestingly, simulations of galaxies at $z\sim2$ suggest that this dichotomous behaviour is to be expected, as the typical sensitivity and beam size of submm surveys (e.g.\ SHADES, LESS; $\sigma_{850}\sim1\,$mJy; Mortier et al.\ 2005; Weiss et al.\ 2009) is such that they should be sensitive to both merger-induced starbursts and the massive end of the star-forming disk population (Hayward et al.\ 2012).

Thus, while SMGs may not form a single homogenous population, submm surveys are able to trace the most highly star forming galaxies at $z\sim2$ in a way not currently possible at other wavelengths. With the advent of {\it Herschel}\footnote{Herschel is an ESA space observatory with science instruments provided by European-led Principal Investigator consortia and with important participation from NASA}(Pilbratt et al. 2010), large scale ($\gs100$\,deg.$^2$) multi-band submm surveys have become possible for the first time (e.g HerMES\footnote{hermes.sussex.ac.uk}, H-ATLAS; Oliver et al.\ 2012; Eales et al.\ 2010). The scientific yield from these surveys has been revolutionary in two ways. Firstly, the massive increase in survey volume (from increases in both depth and area) over existing submm datasets has produced a significant leap in our knowledge of the aggregate properties of the submm population, e.g. their number density (Oliver et al.\ 2010; Clements et al.\ 2010; Glenn et al.\ 2010; B{\'e}thermin et al.\ 2011, 2012a) and clustering properties (Cooray et al.\ 2010; Maddox et al.\ 2010; Amblard et al.\ 2011; Viero et al.\ 2012; van Kampen et al.\ 2012). Secondly, the multi-band nature of the {\it Herschel} data, with six far-IR/submm bands covering the observed frame 70--500$\,\mu$m range, has meant that the physical information contained in the far-IR spectral energy distribution (SED) can be probed for the first time (e.g. Elbaz et al.\ 2010; Hwang et al.\ 2010; Magdis et al.\ 2010; Smith et al.\ 2012a; Symeonidis et al.\ 2013). 

However, the interpretation of the {\it Herschel} data sets has been hampered by the large beam size of {\it Herschel} at submm wavelengths ($\gs20$ arcsec). The resulting confusion noise ($\sim$6\,mJy 1$\sigma$; Nguyen et al.\ 2010) for SPIRE imaging has meant that for the faintest SPIRE sources (i.e. $<30\,$mJy) both accurate cross-identifications and flux densities require the use of ancillary data at IR or radio wavelengths (e.g. Roseboom et al.\ 2010; 2012a). This dependence on ancillary data has meant that the {\it Herschel} surveys have struggled to significantly increase the observable parameter space, in terms of $L_{\rm IR}$--$z$, for IR galaxies.

 While ALMA will eventually produce deep, high resolution, blank field submm imaging, its limited field of view ($\ll1$ arcmin at 450$\,\mu$m) means that even moderately wide-area surveys (e.g. $>100$\,arcmin.$^2$) would be very observationally expensive. For this reason, SCUBA-2 on the 15-m JCMT occupies a unique niche in submm imaging capability (Holland et al.\ 2013). With a mapping speed more than $100\times$ faster than the original SCUBA and a beam size of $8$ arcsec at 450$\,\mu$m, SCUBA-2 offers for the first time both the sensitivity and angular resolution required to perform sensitive studies of individual galaxies at $z>1$ without concerns about source confusion.

The SCUBA-2 Cosmology Legacy Survey (S2CLS) is the largest programme that will be undertaken with SCUBA-2, aiming to map $0.6\ (\sim10)$ deg.$^2$ of the best extragalactic survey fields at 450 (850) $\,\mu$m to a depth of $\sigma\sim 1\,$mJy. Twelve months of S2CLS observing, equating to over $100$\,hrs on-sky, have now been obtained, with the first results on the deep number counts at 450$\,\mu$m presented by Geach et al.\ (2013). Here, we investigate the multi-wavelength properties of the galaxies selected at 450$\,\mu$m using preliminary data in two deep S2CLS 450$\,\mu$m fields; COSMOS and UDS. 

In \S\ref{sec:data} we introduce the data sets utilised in this work and \S\ref{sec:ids} describes the process undertaken to find multi-wavelength identifications for the SCUBA-2 sources. In \S\ref{sec:results} we present our results: the identification statistics and demographics for our 450$\,\mu$m sample (\S\ref{sec:idstats}) and a comparison to measurements of the same sources with {\it Herschel} SPIRE (\S\ref{sec:hercomp}). The scientific implications of these results are investigated via a detailed look at submm SEDs (\S\ref{sec:seds} and \S\ref{sec:smbeta}), and an investigation into the stellar masses and specific SFRs of 450$\,\mu$m sources (\S\ref{sec:ssfr}). Finally, we present our conclusions in \S\ref{sec:conc}.
Throughout we assume a $\Lambda$CDM cosmology with
$\Omega_{\Lambda}=0.7$, $\Omega_{\rm m}=0.3$ and
$H_0=70$\,km\,s$^{-1}$\,Mpc$^{-1}$. All SFR and stellar mass estimates presented assume, or have been converted to, a Chabrier (2003) IMF.

\section{Data}\label{sec:data}
\subsection{SCUBA-2}
The starting point for this study is the 450 and 850$\,\mu$m imaging taken as part of the S2CLS in the deep extragalactic COSMOS and UDS fields. Observations were taken with SCUBA-2 on the 15-m JCMT between October 2011 and July 2012. For these deep fields only the best observing conditions were used, i.e. $\tau_{225\rm GHz}<0.05$. The standard ``daisy'' mapping pattern (Holland et al.\ 2013) was used, with observations broken into 30\,min scans. All data are reduced using the {\sc smurf} package\footnote{http://www.starlink.ac.uk/docs/sun258.htx/sun258.html} V1.4.0 (Chapin et al.\ 2013). The default ``blank field'' configuration was used, although with the Fourier filtering increased to 1.3\,Hz (equivalent to 120 arcsec at the SCUBA-2 scan rate). Flux calibration between pW and Jy is performed using the SCUBA-2 flux calibration factors (FCFs) appropriate for this version of {\sc SMURF}; 606\,Jy\,pW$^{-1}$\,Beam$^{-1}$ for 450\,$\mu$m and 556\,Jy\,pW$^{-1}$\,Beam$^{-1}$ for 850\,$\mu$m (Dempsey et al.\ 2013)\footnote{http://www.jach.hawaii.edu/JCMT/continuum/scuba2/scuba2\_relations.html}. Table \ref{tab:clsobs} details the number of scans taken in each field and the corresponding depths. Noise-only maps of our SCUBA-2 images are produced from the data themselves, by inverting a random half of the individual scans before combining (Austermann et al.\ 2010; Weiss et al.\ 2011). These noise-only maps are used to estimate the completeness and reliability for our source detection algorithm.

\begin{table}
\caption{Details of deep SCUBA-2 CLS fields used in this paper.}
\label{tab:clsobs}
\begin{tabular}{lrrrrrr}
\hline
Field & RA & Dec. & $N_{\rm scans}$ & $\sigma_{450}^{1}$ & $\sigma_{850}^{1}$ & $N(>4\sigma)$\\
\hline
 & deg. & deg. & & mJy & mJy & \\
\hline
COSMOS & 150.124 & 2.29 & 114 & 1.2 & 0.18 & 57\\
UDS &  34.375 & $-5.2$ & 29 & 2.3 & 0.32 & 12\\
\hline
\end{tabular}
$^{1}$ For a point source estimated from exposure time in the deepest region of the map, and not including any contribution from confusion.\\
\end{table}

Point sources are identified in the combined SCUBA-2 450$\,\mu$m images by convolving the map with a ``matched filter'' and looking for peaks which have a signal-to-noise ratio (SNR) of better than four. The matched filter is constructed by first estimating the background on large scales by convolving the map with a Gaussian of ${\rm FWHM}=30$ arcsec. This background is then subtracted from the original map, and the background-subtracted image is convolved with the effective point-source response function (PRF); a Gaussian of ${\rm FWHM}=8$ arcsec which has been ``background--subtracted'' in the same way (see discussion in \S4.2 of Chapin et al.\ 2013).

A total of 69 sources are identified across the two fields above a SNR of four. The SNR threshold of four is used as this is found to produce a false-positive detection rate of 5 per cent, which is considered acceptable for the purposes of this work. Only regions with good coverage (i.e. $\sigma_{450}<5$\,mJy) are used, limiting the area considered in COSMOS and UDS to the central 142 and 77\,arcmin$^{2}$, respectively. Table \ref{tab:clsobs} details the breakdown of these sources across the two fields and their typical noise properties, while Geach et al.\ (2013) presents the number counts derived from the sources detected in the COSMOS field.

Photometry at 850$\,\mu$m is performed simultaneously by assuming the detected 450$\,\mu$m positions and the PRF at 850$\,\mu$m, assumed to be a Gaussian with ${\rm FWHM}=14.6$\,arcsec. The confusion noise at 850$\,\mu$m for SCUBA-2 is predicted to be 0.3\,mJy (Bethermin et al.\ 2011) and so our images will be confusion noise limited in the deepest regions. To account for this we add this addition confusion noise to our instrumental noise estimates in quadrature. A full list of the 450$\,\mu$m sources detected in the COSMOS and UDS fields is given in Tables \ref{tab:s2cosmos} and \ref{tab:s2uds}, respectively. Analysis and catalogues of these deep 850$\,\mu$m pointings independant of the 450$\,\mu$m data will be present in future work.

The completeness of the 450$\,\mu$m source catalogue is assessed by inserting sources of known flux into the noise-only maps. A total of 10$^4$ test sources are used, split into 10 logarithmically-spaced flux bins between 1 and 60$\,$mJy. In order to not adversely affect the overall map statistics, test sources are injected into, and recovered from, the noise-only maps in groups of 20 at a time, resulting in a total of 2000 simulated maps. Sources are recovered from each of these simulated maps using the same methods as the real maps. The completeness as a function of flux is assessed by taking the ratio of the total number of detected sources to the number injected into the noise-only maps.

\subsection{\textit{ Herschel} SPIRE}
Overlapping {\it Herschel} SPIRE (Griffin et al.\ 2010) data exist for both of our SCUBA-2 CLS fields from the HerMES project (Oliver et al.\ 2012). While the instrumental noise of SPIRE data is similar to the SCUBA-2 data at 450$\,\mu$m (i.e. $\sigma=1$--$2\,$mJy), the confusion noise is almost five times higher ($\sim6$\,mJy; Nguyen et al.\ 2010). Thus to obtain reasonable SPIRE photometry for our 450$\,\mu$m sources we must resort to using the prior-based methods described in Roseboom et al.\ (2010) and Roseboom et al.\ (2012a; henceforth R12). Specifically, we follow the methodology presented in R12, with the minor difference that the 4.5$\,\mu$m positions of 450$\,\mu$m sources (as determined by our identification process in \S\ref{sec:ids}) are considered in conjunction with 24$\,\mu$m sources. This work makes use of SPIRE images produced by HerMES, as described in Levenson et al.\ (2010), with updates in Viero et al.\ (2012). As in R12, weightings are used to ensure that rarer sources are upweighted in comparison to more common ones. As the surface density of 450$\,\mu$m detected sources is about ten times lower than that of 24$\,\mu$m sources, the 450$\,\mu$m sources typically have higher weightings and are hence given preference in highly degenerate situations (e.g. a 450$\,\mu$m and 24$\,\mu$m source separated by less than one pixel in SPIRE imaging). 


\subsection{Other ancillary data and photometric redshifts}
The deep S2CLS fields were chosen due to the large amount of ancillary data available at optical and near-IR wavelengths, in particular the ongoing {\it HST} Cosmic Assembly Near-infrared Deep Extragalactic Legacy Survey (CANDELS)\footnote{http://candels.ucolick.org} which is imaging $0.2$ deg.$^2$ to a depth of $F160W<26.5$\,mag (Grogin et al.\ 2011; Koekemoer et al.\ 2011). In both COSMOS and UDS we produce new multi-wavelength catalogues by first convolving all available optical/near-IR imaging to a common PSF ($0.8$ arcsec FWHM) and then performing aperture photometry using the CANDELS $F160W$ as the detection band (see Bowler et al. 2012 for details). For each source in these catalogues we calculate photometric redshifts and stellar masses using the approach presented in Cirasuolo et al. (2010) and Micha{\l}owski et al.\ (2012), which we briefly summarise here. Optical near-IR photometry is compared to a representative grid of galaxy templates, generated from Bruzual and Charlot (2003) models. Small adjustments to the zero-points of the photometric bands are made by testing the photo-$z$ algorithm against available spectroscopic redshifts in both fields (i.e zCOSMOS, UDSz; Lilly et al.\ 2007; Almaini et al., in prep). After adjusting the zero-points, and excluding sources with poor fits to the templates (i.e. those with a chi-squared statistic of $\chi^2>20$), the typical photo-$z$ accuracy is found to be $dz/(1+z)=0.03$. Meanwhile, stellar masses are estimated assuming a two-component stellar population; an old population and a recent "burst".  As discussed by Micha{\l}owski et al.\ (2012), the use of these two-component models to infer the mass of submm-luminous galaxies results in a $\sim0.3$\,dex uncertainty in the mass estimates. 


{\it Spitzer} IRAC 3.6 and 4.5$\,\mu$m photometry is added by directly fitting the IRAC images using $K$-band positions and shape information (UKIDSS in UDS, UltraVISTA in COSMOS). The 5.8 and 8$\,\mu$m IRAC channels are not considered as they suffer from source blending issues due to the large IRAC PSF at these wavelengths. Across most of the redshift range of interest these channels do not probe the stellar light from galaxies, and hence their omission does not significantly affect photo-$z$ and stellar mass estimates. 

{\it Spitzer} MIPS 24$\,\mu$m photometry in COSMOS is performed using the {\sc starfinder} package (Diolaiti et al. 2000) on the public SCOSMOS image (Le Floc\'{h} et al.\ 2009), as described in Roseboom et al.\ (2012b).



\section{Identifications for SCUBA-2 sources}\label{sec:ids}
Despite the relatively small beam of SCUBA-2 at 450$\,\mu$m, the low signal-to-noise ratio (SNR) nature of our source catalogues ($>4\sigma$), as well as the matching depth at optical/near-IR wavelengths required to find identifications ($K_{\rm AB}\sim24$; Oliver et al.\ 2012), means that simple nearest-neighbour matching is not sufficient. Thus we make use of the likelihood ratio (LR) formalism, as presented by Sutherland \& Saunders et al. (1992), to make statistically reliable cross-identifications between our SCUBA-2 450$\,\mu$m source lists and the ancillary data. 

In order to utilise both colour and mid-IR information in the matching process we extend the LR method in a similar fashion as in Chapin et al.\ (2011; henceforth C11). Specifically, for each match between a 450$\,\mu$m source $i$ and ancillary source $j$ within a maximum separation $r_{\rm max}$ we calculate:
\begin{equation}
L_{i,j}=\frac{q(S,c)e^{-r^2_{i,j}/2\sigma_{i}^2}}{2\pi\sigma_{i}^2\rho(S,c)},
\label{eqn:lr}
\end{equation}
where $r_{i,j}$ is the separation between the two sources, $\sigma_i$ is the positional uncertainty of the SCUBA-2 450$\,\mu$m source, $\rho(S,c)$ is the probability of finding a background source with the flux density $S$ and a colour $c$, while similarly $q(S,c)$ is the prior probability of a true match having this flux and IRAC colour. For this work the matching flux density $S$ is the 24$\,\mu$m value where $S_{24}>80\,\mu$Jy ($4\sigma$ for the fields considered here), otherwise the 4.5$\,\mu$m value is used. The 24\,$\mu$m identifications are generally preferred as the areal density of 24$\,\mu$m sources is significantly lower than 4.5$\,\mu$m sources ($\sim10\times$), and the 24$\,\mu$m photometry is typically dominated by thermal dust emission as opposed to starlight which dominates at 4.5$\,\mu$m. While both bands could have been used (as in C11), we find $S_{4.5}$ and $S_{24}$ to be correlated (a Spearman rank test gives a correlation coefficient of 0.6), and hence there is little extra discriminatory power in using both bands. For the colour $c$ we use the colour between the two IRAC channels at 3.6$\,\mu$m and 4.5$\,\mu$m: $[3.6-4.5]$.

For the positional uncertainty, $\sigma_i$, we assume $\sigma_i=0.6\,{\rm FWHM}/{\rm SNR}$ (Ivison et al.\ 2007), with a minimum uncertainty of 1 arcsec to account for the overall pointing accuracy of JCMT. 

The prior probabilities  $q(S,c)$ and $\rho(S,c)$ are calculated directly from the data in a similar fashion to C11. For each source in our 450$\,\mu$m catalogue we find all of the potential matches in the IRAC and 24$\,\mu$m catalogues within a maximum search radius of 4 arcsec. This radius was chosen as it is $\sim\frac{1}{2}$\,FWHM, i.e. $3\sigma_i$ for SNR$=4$ sources. As in C11 we assume the IR flux and IRAC colour are independent and so $q(S,c)\approx q(S)q(c)$ and similarly $\rho(S,c)\approx\rho(S)\rho(c)$. This assumption is justified, as little correlation is seen between $S_{24}$ or $S_{4.5}$ and $[4.5-3.6]$ colour, with a Spearman rank test giving correlation coefficients of 0.1 and 0.03, respectively.

\begin{figure*}
\includegraphics[scale=0.6]{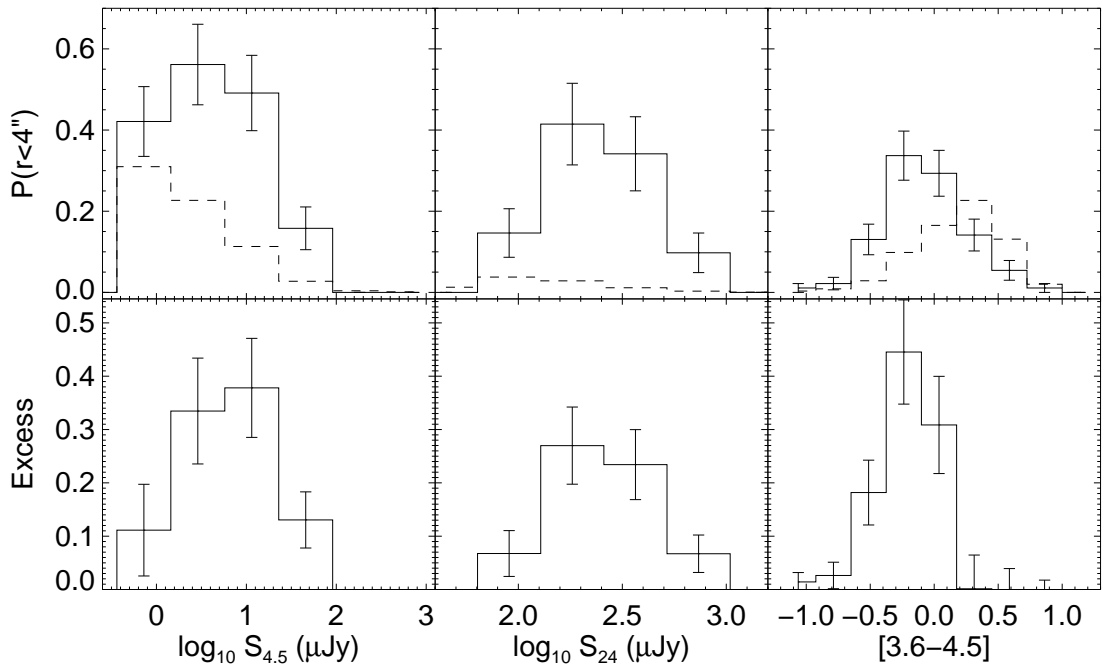}
\caption{Statistical excess of matching sources as a function of $S_{4.5}, S_{24}$ and $[3.6-4.5]$ colour. Top panels: probability of finding a source as a function of $S_{4.5}$ (left), $S_{24}$ (middle) and $[3.6-4.5]$ (right), within 4 arcsec of a SCUBA-2 450$\,\mu$m source (solid line) or in a random 4 arcsec radius aperture (dashed line). Bottom panels: excess probability of finding a source within 4 arcsec of a 450$\,\mu$m position.}
\label{fig:lrqp}
\end{figure*}

Figure \ref{fig:lrqp} shows the probability of finding a source as a function of $S_{4.5}, S_{24}$ and $[4.5-3.6]$ within 4 arcsec of a SCUBA-2 450$\,\mu$m source compared to a random 4 arcsec radius aperture. A clear excess at both IR fluxes and $[3.6-4.5]$ colour can be seen. The integral $q_0=\int q(S)dS$ gives an estimate of the fraction of counterparts to 450$\,\mu$m sources we should expect to recover. For the $S_{4.5}$, $S_{24}$ and $[3.6-4.5]$ excess we find $q_0=0.96, 0.64$ and $0.98$, respectively. Thus we would expect almost all of our 450$\,\mu$m sources to have counterparts above the depth of the matching catalogue ($F160W<26.5$), and $64$ percent of these to also have 24$\,\mu$m detections. However, it is worth noting that this assumes that each 450$\,\mu$m source has a single counterpart, i.e. the 450$\,\mu$m flux originates from a single, unique, source at 4.5$\,\mu$m.

In order to determine the false-positive rate of cross-identifications we perform Monte Carlo simulations of the matching process. We generate 10,000 mock source positions at random locations within the SCUBA-2 450$\,\mu$m field. For each random position the positional uncertainty is again calculated from the SNR at 450$\,\mu$m, with the value drawn randomly from the real distribution of observed values. Using the same prior distributions (i.e. $q(S)$ and $q(c)$) as for the real data, we attempt to find identifications in the matching catalogue for these mock sources and calculate the likelihood ratio of each match. 

Using these simulations we evaluate the false positive rate by taking the ratio of the number of matches in the simulation to the total number of random positions.

Interpolating these simulations, an association has a 20 per cent chance of being a spurious at $\log_{10}L=1.1$. In the COSMOS and UDS fields 49 and 9 450$\,\mu$m sources, respectively, have at least one counterpart in the matching catalogues with separation less than 4 arcsec and $\log_{10}L>1.1$. The vast majority of these have $\log_{10}L\gg1.1$; taking our measured values of $\log_{10}L$ for each identification, and using our simulations as a guide, we expect 2.3 of our combined total of 58 identifications (four percent) to be spurious. 

Multiple reliable identifications are found for eleven sources (19 per cent).  Given the high reliability of these alternative identifications it is likely that in these cases the submm flux density is split between the identifications, i.e. both identifications are submm bright and simply blended by the 8 arcsec SCUBA-2 beam. Interestingly the rate of multiple identifications is similar to the blending rate found for 850$\,\mu$m selected samples (e.g. Wang et al.\ 2011; Karim et al.\ 2013). For simplicity, we consider the source with the best $\log_{10}L$ value to be the identification and discard the second identification. These alternative identifications are listed in Tables \ref{tab:s2cosmos} and \ref{tab:s2uds}.




\section{Results}\label{sec:results}
\subsection{Identification statistics and demographics of the 450$\,\mu$m population}\label{sec:idstats}

The identification rate of 450$\,\mu$m sources is $86\pm12$ and $75\pm25$ per cent for the COSMOS and UDS fields, respectively. This leads to a combined identification rate of $84\pm11$ per cent. When compared with our estimate of the potential identification rate from the excess of sources around 450$\,\mu$m sources, i.e. $q_0=0.98$, there is a deficit of $14$ per cent, i.e. there are potentially eleven sources which have counterparts in our matching catalogues that could not be reliably recovered by our identification process. Indeed, all of these sources have potential identifications within the maximum search radius (4 arcsec), but are deemed unreliable. However, this deficit may also highlight a weakness in the likelihood ratio methodology, as it assumes no clustering of the matching catalogue around the 450$\,\mu$m sources. It is well known that this is not true, in fact submm sources are strongly clustered (e.g. Viero et al.\ 2012; Karim et al.\ 2013) and so $q_0$ will always be biased. 

\begin{figure}
\includegraphics[scale=0.4]{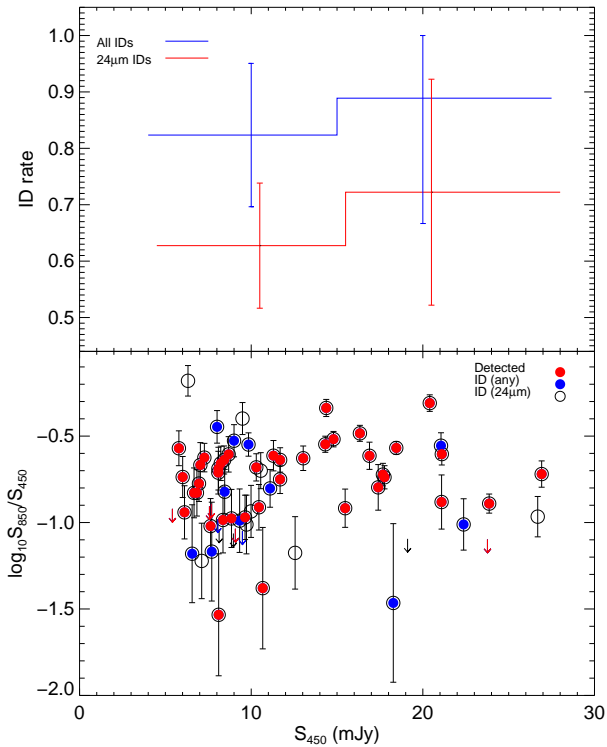}
\caption{Top: ID rate as a function of 450$\,\mu$m flux density. The identification rate is broadly insensitive to $S_{450}$, with a near constant level of about $80$ per cent for all identifications, and $60$ per cent for identifications with $S_{24}>80\,\mu$Jy. Bottom: Submm colour ($S_{850}/S_{450}$) vs. 450$\,\mu$m flux density. Open symbols are all catalogued 450$\,\mu$m sources, those with reliable identifications are shown as filled symbols (blue for all identified, red for 24$\,\mu$m). Identifications appear insensitive to both flux density and submm colour, suggesting that our sample is unbiased in terms of redshift and luminosity. }
\label{fig:fid}
\end{figure}

If we consider only sources with 24$\,\mu$m identifications above the limits of the surveys considered here (i.e. $S_{24}>80\,\mu$Jy) the rate drops to $65\pm10$ per cent, in good agreement with our estimate from excess sources: $q_0^{24}=0.64$. This number is important as it is common for {\it Herschel} photometry to be performed using the prior positions of sources at $24\,\mu$m as a guide (e.g. Roseboom et al.\ 2010). While this approach may be appropriate for brighter flux densities (i.e. $S_{450}>20\,$mJy), or shorter submm wavelengths than we consider here, it is clear from Fig.~\ref{fig:fid} at $S_{450}\sim10$\,mJy almost half of the submm sources do not have $24\,\mu$m counterparts to a depth of $S_{24}\ge80\,\mu$Jy.

Figure \ref{fig:fid} considers the identification rate as a function of both $S_{450}$ and the submm colour, $S_{850}/S_{450}$. Interestingly, the identification rate is only modestly dependent on source flux density, with the overall rate dropping from 87$\pm24$ per cent at $S_{450}>15\,$mJy to $82\pm13$ per cent at $5<S_{450}<15\,$mJy. The effect for 24$\,\mu$m identifications is larger, falling 11 per cent across the same range. 

Only a weak relationship between submm colour and $S_{450}$ is seen in Fig.~\ref{fig:fid}, with the faintest sources showing some preference for bluer submm colours (lower $S_{850}/S_{450}$). Interestingly, the eleven 450$\,\mu$m sources without reliable identifications do not show any preference in submm colour. If we consider the colour $S_{850}/S_{450}$ to be a crude tracer of redshift, and combined with the weak dependance of the identification rate on $S_{450}$, this suggests that our ability to obtain optical/near-IR identifications for our 450$\,\mu$m sources is independent of both redshift and luminosity.

\begin{figure}
\includegraphics[scale=0.35]{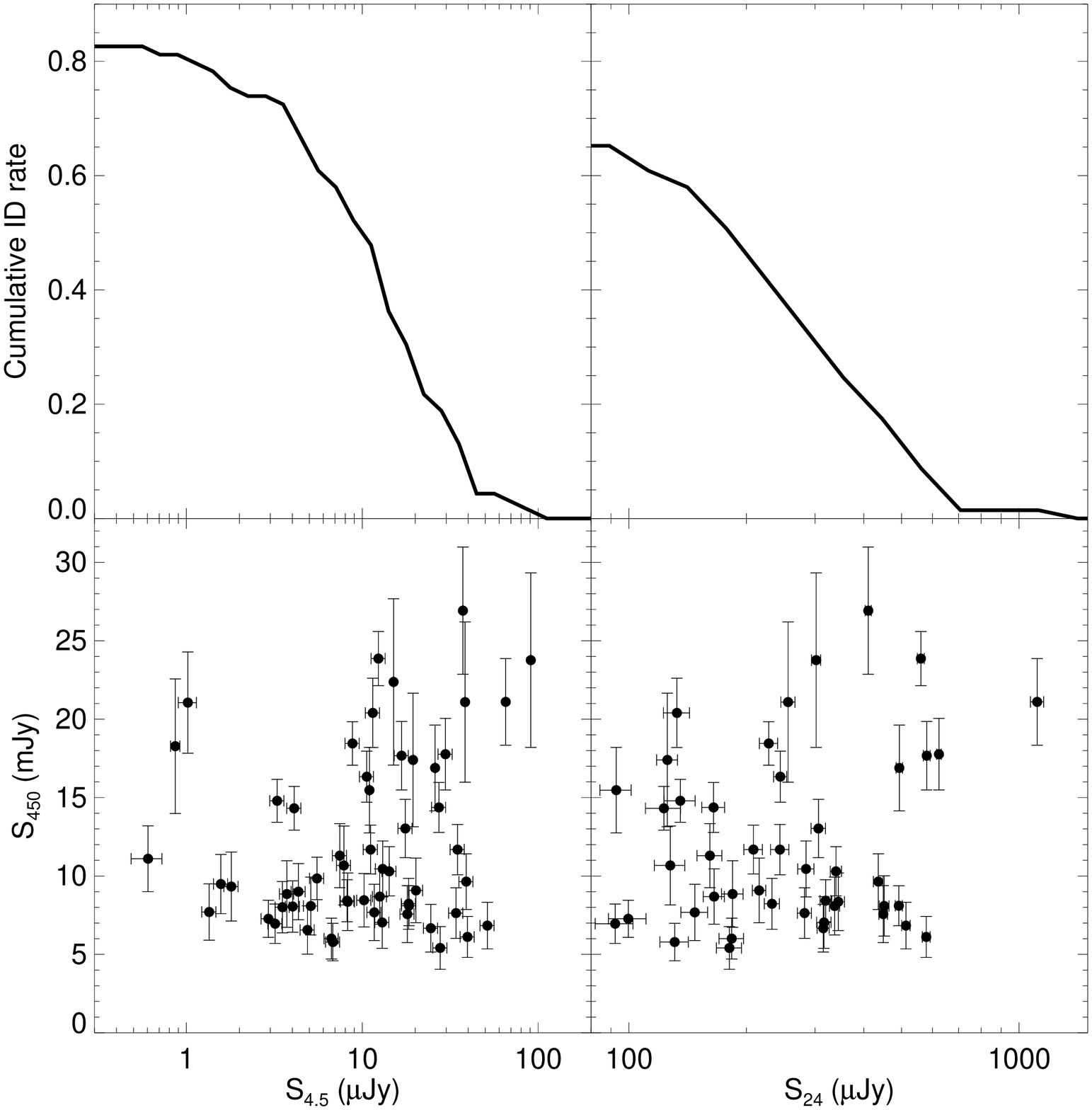}
\caption{Top panels: Cumulative identification rate as a function of decreasing depth for 4.5$\,\mu$m (left) and 24$\,\mu$m (right). Bottom panels: matching flux density vs. $S_{450}$ for 4.5$\,\mu$m (left) and 24$\,\mu$m (right). Both 4.5$\,\mu$m and 24$\,\mu$m flux density are seen to be weakly correlated with 450$\,\mu$m flux density.}
\label{fig:ff}
\end{figure}

In Fig.~\ref{fig:ff} we consider the cumulative identification rate as a function of decreasing flux density at 4.5 and 24$\,\mu$m, i.e. how deep do ancillary data need to be to return identifications for 450$\,\mu$m-selected sources? From Fig.~\ref{fig:ff} we estimate that to return $50$ per cent of the identifications would require a depth of $S_{4.5}=10\,\mu$Jy and $S_{24}=180\,\mu$Jy. In the bottom panels of Fig.~\ref{fig:ff} we show the relationship between $S_{4.5}$, $S_{24}$ and $S_{450}$. In each case a weak correlation can be seen, with the brightest 450$\,\mu$m sources tending to also be bright at 4.5 and 24$\,\mu$m. 

The weak relationship between 450$\,\mu$m and 4.5/24$\,\mu$m flux density is intriguing, and potentially worrying, as it may point to a bias away from identifications for the sources with the faintest counterparts. To confirm that the depth of the CANDELS-SCOSMOS/SpUDS data combination is not responsible for either the failure to acquire identifications nor a bias towards bright, but erroneous, identifications we visually inspect the region around each SCUBA-2 sources in the deep {\it Spitzer} Extended Deep Survey (SEDS) 4.5$\,\mu$m images (Ashby et al.\ 2013). These images have a completeness of $\sim70$ percent at ABmag$=24.25$ ($0.7\,\mu$Jy), more three times deeper than IRAC data used here. In none of the 69 sources is there an IRAC source below our current detection limit which is either closer than the current identification, or would have a larger LR value. From this analysis we conclude that our current list of identification is robust, and the 14 sources without identification must either be spurious, or simply nightmare cases for the LR methodology (i.e. larger than expected positional offsets, unusual IRAC colours, etc).

 \begin{figure}
\includegraphics[scale=0.35]{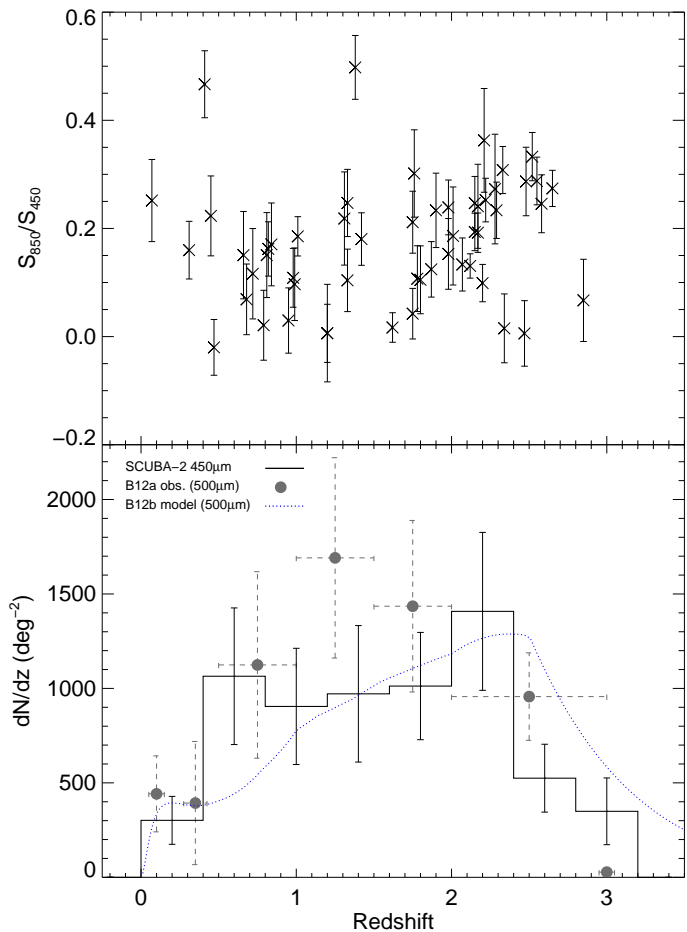}
\caption{Top: SCUBA-2 submm colour ($S_{850}/S_{450}$) vs. redshift for sources with good identifications and optical/near-IR photometric redshifts. Bottom: Redshift distribution ($dN/dz$) of 450$\,\mu$m sources. The solid line is the $dN/dz$ including corrections for completeness in terms of both 450$\,\mu$m detection, and the identification, process. Errors are estimated assuming Poisson statistics. Overplotted as grey points is the equivalent estimate of the $dN/dz$ at SPIRE 500$\,\mu$m achieved via stacking by B{\'e}thermin et al.\ (2012a; B12a). Also shown is the prediction from the latest evolution model from B{\'e}thermin et al.\ (2012b; B12b). Good agreement is seen between the SCUBA-2 450$\,\mu$m and both the model and SPIRE 500$\,\mu$m stacking results.}
\label{fig:zhist}
\end{figure}

Armed with robust counterparts for more than $80$ per cent of our sample we can investigate the demographics of the 450$\,\mu$m population. While 58 sources across UDS and COSMOS have reliable identifications, good photometric redshift estimates are only possible for 54 of these; in four cases the fit to the optical/near-IR photometry is too poor to obtain a reliable photometric redshift estimate (i.e. photo-$z$ $\chi^2>20$). Fig.~\ref{fig:zhist} compares the optical/near-IR photometric redshifts for these 54 sources to the submm colour ($S_{850}/S_{450}$). A weak correlation can be seen, with a Spearman rank correlation coefficient of 0.3 ($p=0.03$ for no correlation). Also shown in Fig.~\ref{fig:zhist} is the redshift distribution ($dN/dz$) for 450$\,\mu$m sources with $S_{450}>6\,$mJy. To estimate the true $dN/dz$ two corrections need to be applied to the observed redshift distribution; 1) a correction for the completeness (and variable depth) of the 450$\,\mu$m data; and 2) a correction for the incompleteness in identifications/redshifts. The first correction is calculated using the completeness and area estimates from Section \ref{sec:data} specifically, in a given redshift bin the redshift distribution is calculated via $dN/dz=\sum_i1/ca$, where $c$ is the completeness for source $i$ and $a$ is the combined area of the COSMOS and UDS fields. To estimate the redshift distribution of sources without reliable redshift information we use the weak correlation observed between redshift and $S_{850}/S_{450}$. For each source without a redshift or identification we build a probability distribution $p(z)$ using the redshifts and colours of the identified sample as a guide, i.e. for a source $i$,
\[
p_i(z)=\frac{1}{N\pi\sigma_c}\sum_j^{N(z)}\exp\left(-\frac{(c_i-c_j)}{2\sigma_c^2}\right),
\]
where the summation runs over the $N(z)$ sources $j$ that fall into a redshift bin centred on $z$, $c_i$ and $c_j$ are the submm colours of the sources, and $\sigma_c=0.1$ (the typical error in the submm colour). The submm colours of the non-identified sources are very similar to those with identifications and so the resulting correction to the observed $dN/dz$ is quite flat, i.e. similar to simply multiplying the observed $dN/dz$ by a constant factor at all redshifts.

The redshift distribution of 450$\,\mu$m sources is broad, peaking in the range $0.5<z<2.5$. Few sources are seen at low-$z$ ($z<0.5$), and high-$z$ ($z>3$). The overall median of the observed $dN/dz$ is $z=1.4$, in good agreement with previous work on this dataset by Geach et al.\ 2013, although somewhat lower than the mean of $z=2$ found for the more luminous 450$\,\mu$m sample presented in Casey et al.\ (2013). Reasonable agreement, both in terms of the normalisation and shape, is seen between our estimate and both the observed and predicted redshift distribution of SPIRE 500$\,\mu$m sources from B{\'e}thermin et al. (2012a, 2012b). Interestingly, the shape of our $dN/dz$ more closely resembles the B{\'e}thermin et al. (2012b; B12b) model prediction, broad with a peak close to $z=2.2$, than the B{\'e}thermin et al.\ (2012a; B12a) result from SPIRE data, which peaks more prominently at $z\sim1.3$. 

We can quantify these comparisons via the use of Kolgmogorov-Smirnov (KS) tests. In the case of the B12b model prediction we can test directly our SCUBA-2 sample directly, by considering the B12b prediction taking into account the completeness as a function of redshift for our SCUBA-2 catalogue. A KS test suggests a 88 per cent probability that our 450$\,\mu$m sample is drawn from the B12b model distribution. For the B12a observational result it is more difficult to apply this type of testing, as the result is derived not from redshifts of individual sources, but from stacking in bins of flux density. Nonetheless we compare the B12a result to our sample by generating Monte-Carlo realisations of the B12a results, taking into account the quoted errors. In each realisation we draw 10,000 galaxies from the B12a distribution and apply the redshift dependant completeness of our sample. After generating 10,000 realisations we find that in 40 per cent of realisations a KS test would not distinguish between the two distributions (defined as $p_{\rm KS}>0.05$). Thus while the B12a and our $dN/dz$ may appear significantly different in Fig.~\ref{fig:zhist}, the significant errors mean that two results are not strictly in conflict.



\subsection{Comparing SCUBA-2 and SPIRE photometry}\label{sec:hercomp}
The increased resolution of SCUBA-2 at 450$\,\mu$m offers the opportunity to assess the impact of the poor angular resolution of SPIRE on photometry at similar submm wavelengths and depths. The SPIRE 500$\,\mu$m waveband is the most directly comparable as it completely encompasses the SCUBA-2 450$\,\mu$m bandpass, although with a resolution of 36 arcsec. 

A variety of methods have been developed to perform photometry at SPIRE wavelengths, which can be broken into two broad classes: those which use prior information at other wavelengths to try to ``de-confuse'' the SPIRE photometry (e.g.\ Roseboom et al.\ 2010, 2012), and those which use traditional point-source extraction methods without considering other information (e.g. SussExtractor: Smith et al.\ 2010). In between these methods are a number of ``hybrid'' methods, which try to build complete lists of SPIRE sources via point source detection across the SPIRE bands, and then perform photometry at the three wavelengths simultaneously using these detected sources (e.g. Wang et al., in prep).

Armed with our high resolution 450$\,\mu$m data we can evaluate the effectiveness of these different approaches. To begin, we cross-match our SCUBA-2 450$\,\mu$m catalogue to three different versions of SPIRE catalogues produced by the HerMES project: The `XID24' catalogues, which use a {\it Spitzer} 24$\,\mu$m prior (Roseboom et al.\ 2012); the `XID250' catalogues, which produce a source list at 250$\,\mu$m and use these sources as a prior at 500$\,\mu$m (Wang et al., in prep); and the SusseXTractor (SXT) catalogues, which simply perform point-source detection and extraction at 500$\,\mu$m (Smith et al.\ 2010). The XID24 catalogues are matched to our SCUBA-2 catalogue using a 4 arcsec matching radius between the 450$\,\mu$m position and the 24$\,\mu$m position used for the XID24 extraction. The XID250 and SXT catalogues are matched to our SCUBA-2 positions using a 10 arcsec search radius ($\sim1/3$ of the beam FWHM for SPIRE at 500$\,\mu$m).

For the 69 SCUBA-2 450$\,\mu$m sources we find: 53 counterparts in the XID24 catalogue, 49 counterparts in the XID250 catalogue and 13 counterparts in the SXT catalogue. For the prior driven catalogues (i.e. XID24 and XID250) the option exists within the source extraction process to set the 500$\,\mu$m flux to zero, i.e. the algorithm suggests that no significant 500$\,\mu$m source is present at this position. Amongst the 53 counterparts in the XID24 catalogue 44 have non-zero 500$\,\mu$m flux densities. For the XID250 counterparts 33 sources have non-zero 500$\,\mu$m flux densities. 

\begin{figure}
\includegraphics[scale=0.45]{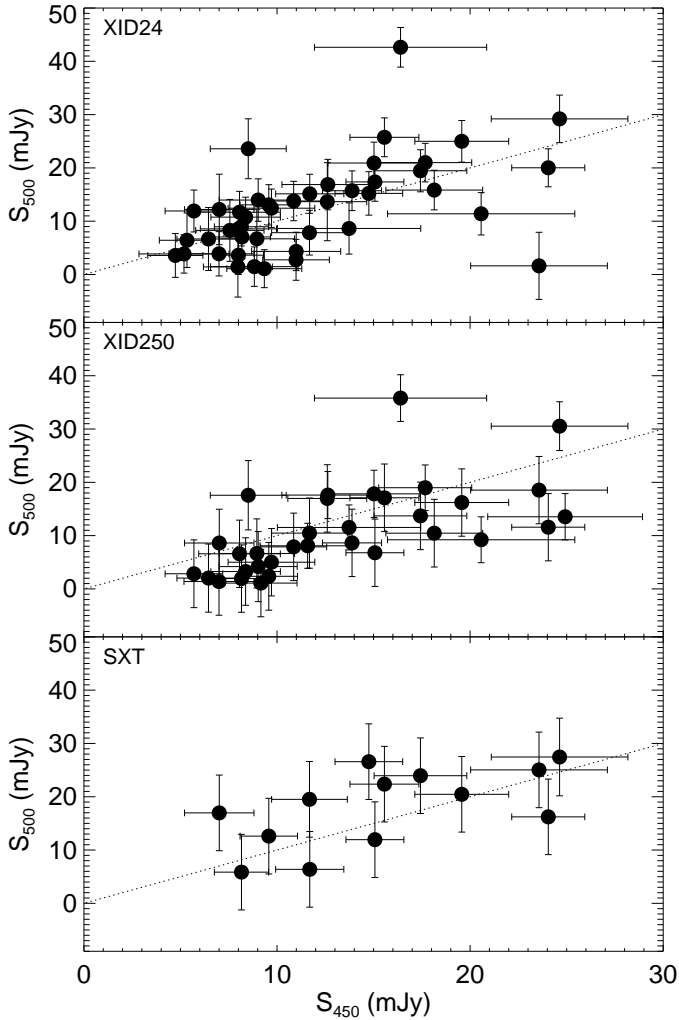}
\caption{Comparison between SCUBA-2 450$\,\mu$m and SPIRE 500$\,\mu$m flux densities for sources in common between our 450$\,\mu$m catalogue and the HerMES XID24, XID250 and SXT catalogues. For the majority of sources the XID24 catalogued fluxes agree well with our SCUBA-2 values. The dotted line in each panel shows the one-to-one correspondence. Both the XID250 and SXT catalogues offer worse agreement, with the XID250 values well below the SCUBA-2 measurements at bright ($\gs 20\,$mJy) values, while the SXT values show a wide scatter around the SCUBA-2 measurements.} 
\label{fig:hermescomp}
\end{figure}

In Fig.~\ref{fig:hermescomp} we compare the SCUBA-2 450$\,\mu$m flux densities to the SPIRE 500$\,\mu$m flux densities for the matched sources in the XID24, XID250 and SXT catalogues. The XID24 flux densities show good agreement with our SCUBA-2 measurements for the majority of sources down to the detection limit for SCUBA-2 ($S_{450}\sim5$\,mJy), with a small number of outliers; only 2 (out of 53; 4 per cent) sources exhibit $|S_{500}-S_{450}|/\Delta S_{500}>3$. This is an impressive achievement, as the confusion limit (40 beams/src) at 500$\,\mu$m for SPIRE is about $15\,$mJy (Nguyen et al.\ 2010), and so it appears that the use of the 24$\,\mu$m prior has the ability to produce reliable flux densities well below this limit. The XID250 flux densities show broad agreement with the SCUBA-2 estimates, although they appear to be systematically lower, particularly at bright 450$\,\mu$m flux density. While SXT flux density estimates only appear for a small number of 450$\,\mu$m sources, they too show broad agreement, albeit with a large scatter.
\section{Discussion}\label{sec:disc}%
\subsection{Submm SEDs}\label{sec:seds}
A common way of interpreting submm SEDs is to fit a modified blackbody of the form;
\begin{equation}
S_{\nu}\propto\left\{1-\exp\left[-\left(\frac{\nu}{\nu_0}\right)^{\beta_{\rm D}}\right]\right\}B_{\nu}(T_{\rm D}),
\label{eqn:mbb}
\end{equation}

\noindent where $T_{\rm D}$ is the dust temperature, $\nu_0$ the rest-frame frequency at which the emission becomes optically thick, $\beta_{\rm D}$ the emissivity and $B_{\nu}$ the Planck function (e.g.\ Blain et al.\ 2003; Roseboom et al.\ 2012a; Magnelli et al.\ 2012). It is also common to replace the Wien side of the blackbody SED with a power-law at some arbitrary cut-off, i.e. $S_{\nu}\propto\nu^{-\alpha}$ for $\nu$, where $\frac{\partial S}{\partial\nu}\ls-\alpha$. While this type of SED fitting is not very physically motivated, for submm datasets with limited wavelength sampling it offers an effective mechanism for interpolating between the data points (enabling the integrated IR luminosity to be determined), as well as a small set of metrics that crudely describe the shape of the submm SED.

Unfortunately, even this simple parameterisation of the SED can be difficult to constrain with typical submm datasets, and so a number of the parameters (generally the emissivity, $\beta_{\rm D}$, the transition frequency, $\nu_0$, and Wien side slope, $\alpha$) are set to assumed values. This is problematic, as recent studies at low-$z$ are showing that these SED parameters do vary significantly and are a function of galaxy properties (Auld et al.\ 2012, Smith et al.\ 2012, Galametz et al.\ 2012). Despite valiant attempts (Chapin et al.\ 2009; Kovacs et al.\ 2011, Magnelli et al.\ 2012), typical values for distant galaxies, and their relationship with other properties, are not known.

\begin{figure*}
\includegraphics[scale=0.6]{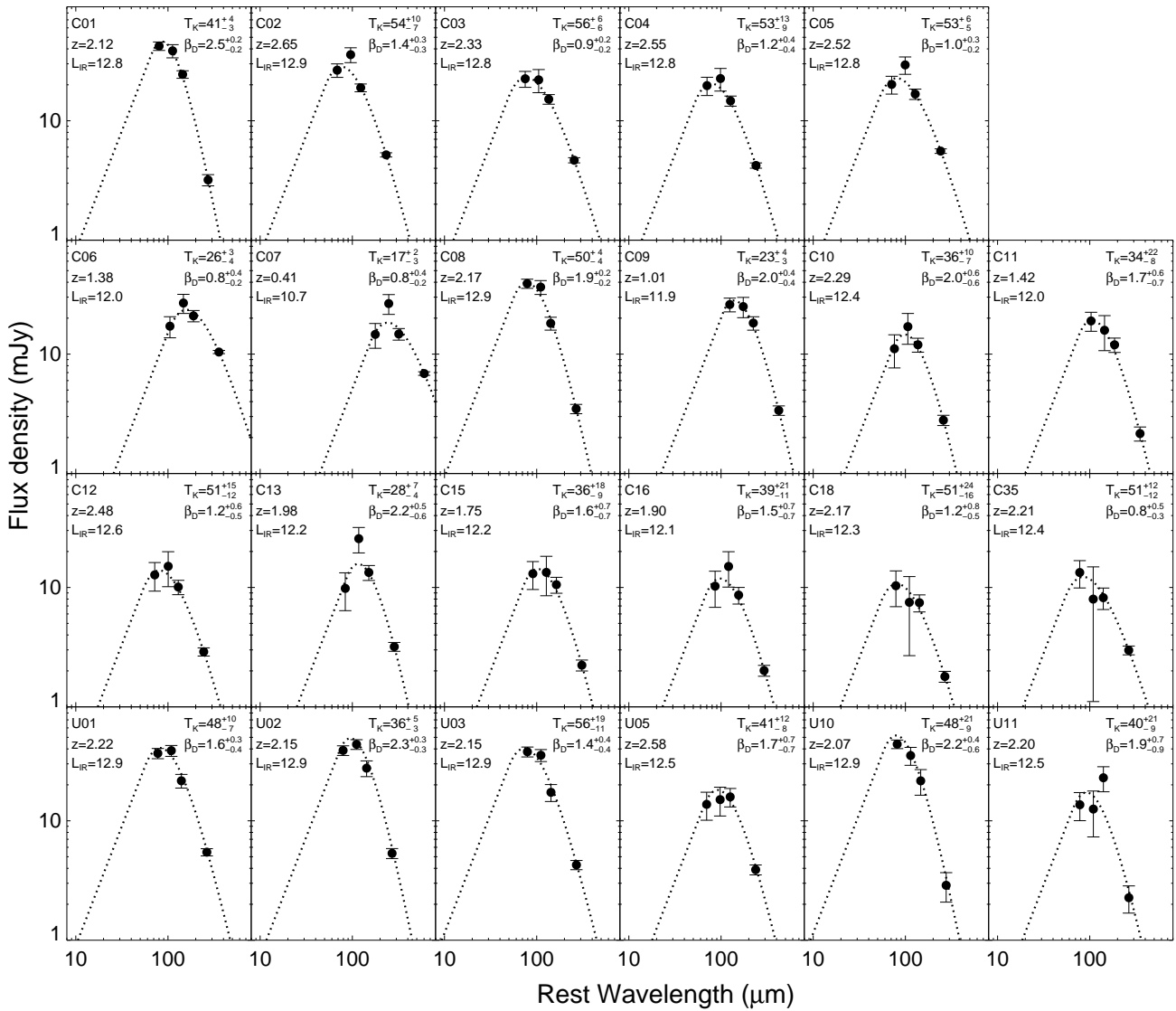}
\caption{Modified blackbody fits to the submm SEDs for the 23 450$\,\mu$m sources which have robust photometry in at least three submm bands. In the majority of cases the modified blackbody offers an excellent description of the submm photometry.}
\label{fig:seds}
\end{figure*}

The shortest observed-frame wavelength considered here is 250$\,\mu$m, which translates to rest-frame 63$\,\mu$m at the maximum redshift of our sample ($z=3$). Thus our ability to constrain parameters sensitive to the Wien side of the SED (e.g. $\nu_0$, $\alpha$) is limited. However, the combination of SPIRE and SCUBA-2 does provide excellent coverage of the Rayleigh-Jeans tail of the SED, and so parameters sensitive to this (e.g. $\beta_{\rm D}$) should be well constrained by our data set.

To begin, we select the 23 450-$\mu$m sources which have robust ($>3\sigma$) photometry in at least three of the four submm wavebands: SPIRE at 250 and 350$\,\mu$m and SCUBA-2 at 450 and 850$\,\mu$m. To these sources we fit modified blackbody curves (Eq.~\ref{eqn:mbb}), allowing $L_{\rm IR}$, $T_{\rm D}$  and $\beta_{\rm D}$ to vary, but holding the transition frequency fixed at a rest-frame value of $\nu_0=c/100\,\mu$m, and the Wien side slope to $\alpha=2$. While these values of $\nu_0$ and $\alpha$ are arbitrary our results are not highly sensitive to it; repeating our SED fitting with a range of $\nu_0$ values between $\nu_0=c/50$--$c/200\,\mu$m, and $\alpha=1.5$--$2.5$, changes our estimates of the other SED parameters typically by $<10$ per cent. The SED fits are performed using an MCMC approach which allows both the best-fit parameter values and variances to be robustly determined. Full details of our MCMC fitting approach are given in Appendix \ref{app:mcmc}.

The requirement of good photometry in three bands is effectively a $S_{450}>11.5$\,mJy selection; 19 out of the 23 sources with good photometry have $S_{450}>11.5$\,mJy, while only two sources with $S_{450}>11.5$\,mJy are excluded. The vast majority of the 34 sources excluded fail due to having unreliable photometry in the {\it Herschel} SPIRE bands, this is unsurprising as 11.5\,mJy is significantly below the confusion limit for SPIRE ($\sim 18$\,mJy, 3$\sigma$; Nguyen et al.\ 2010).

Fig.~\ref{fig:seds} shows the resulting SED fits to the submm photometry. In the majority of cases the submm SED is well-described by the modified blackbody form. In a number cases the SPIRE 350$\,\mu$m photometry point is inconsistent with the modified blackbody, being a bit higher than the expected value from the fit. We assume that residual confusion noise in the SPIRE flux extraction is the cause of this discrepancy, which is included in the flux density noise estimates.

To assess how well our limited submm photometry can constrain the modified blackbody parameters we can use the estimates of the posterior probability distribution for each parameter produced by the MCMC. The typical error on our estimates of the dust temperature and emissivity (averaging the asymmetric errors) is $\Delta T_{\rm D}=10$\,K and $\Delta\beta_{\rm D}=0.4$, respectively, although it is worth noting that these parameters are degenerate, and so the errors are correlated (i.e. increasing $T_{\rm D}$ is degenerate with decreasing $\beta_{\rm D}$; Shetty et al.\ 2009a). For the mean parameter values, $\langle T_{\rm D}\rangle=42$\,K and $\langle\beta_{\rm D}\rangle=1.6$, this translates to an error of $\sim 20$ per cent in both cases. Given this, we can be confident that our submm photometry contains sufficient information to usefully constrain the parameters of the modified blackbody model.

\begin{figure}
\includegraphics[scale=0.5]{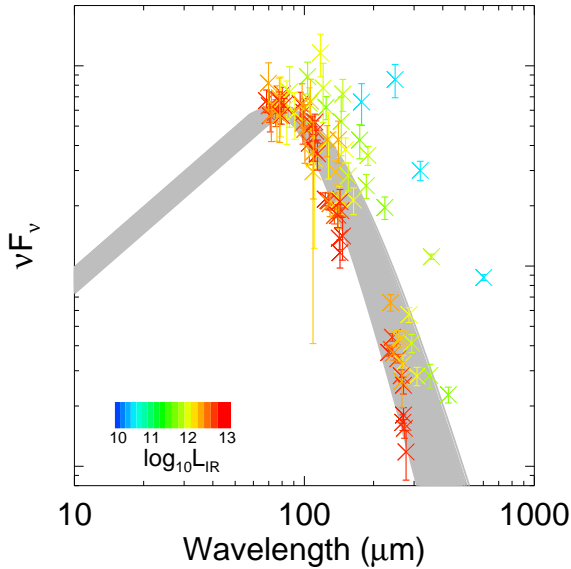}
\caption{Rest-frame, luminosity normalised, submm photometry for the 23 SCUBA-2 450$\,\mu$m sources considered here (crosses). Photometric points are colour-coded by $L_{\rm IR}$. Also shown in grey is the range of modified blackbody SED shapes defined by $T_{\rm D}=42\pm11$\,K and $\beta_{\rm D}=1.6\pm0.5$. With the exception of a few low-luminosity, low-$T_{\rm D}$, sources, the photometry for most 450$\,\mu$m population is consistent with this range of SEDs. }
\label{fig:sedcombo}
\end{figure}

For both the dust temperature and emissivity the variance in best fit parameters is only slightly larger than the typical error, with $\sigma(T_{\rm D})=11$\,K and $\sigma(\beta_{\rm D})=0.5$. This suggests that assuming single values of the SED parameters would provide a reasonably adequate description of submm population; the spread in $T_{\rm D}$ and $\beta_{\rm D}$ is small enough that taking a single SED with $T_{\rm D}=42$\,K and $\beta_{\rm D}=1.6$ would be sufficient to explain the submm photometry for the majority of our sources. This is illustrated in Fig.~\ref{fig:sedcombo}, which compares the rest-frame, luminosity normalised, photometry for these 23 SCUBA-2 sources to the range in modified blackbody SEDs defined by $T_{\rm D}=42\pm11$\,K and $\beta_{\rm D}=1.6\pm0.5$.



\begin{figure*}
\includegraphics[scale=0.6]{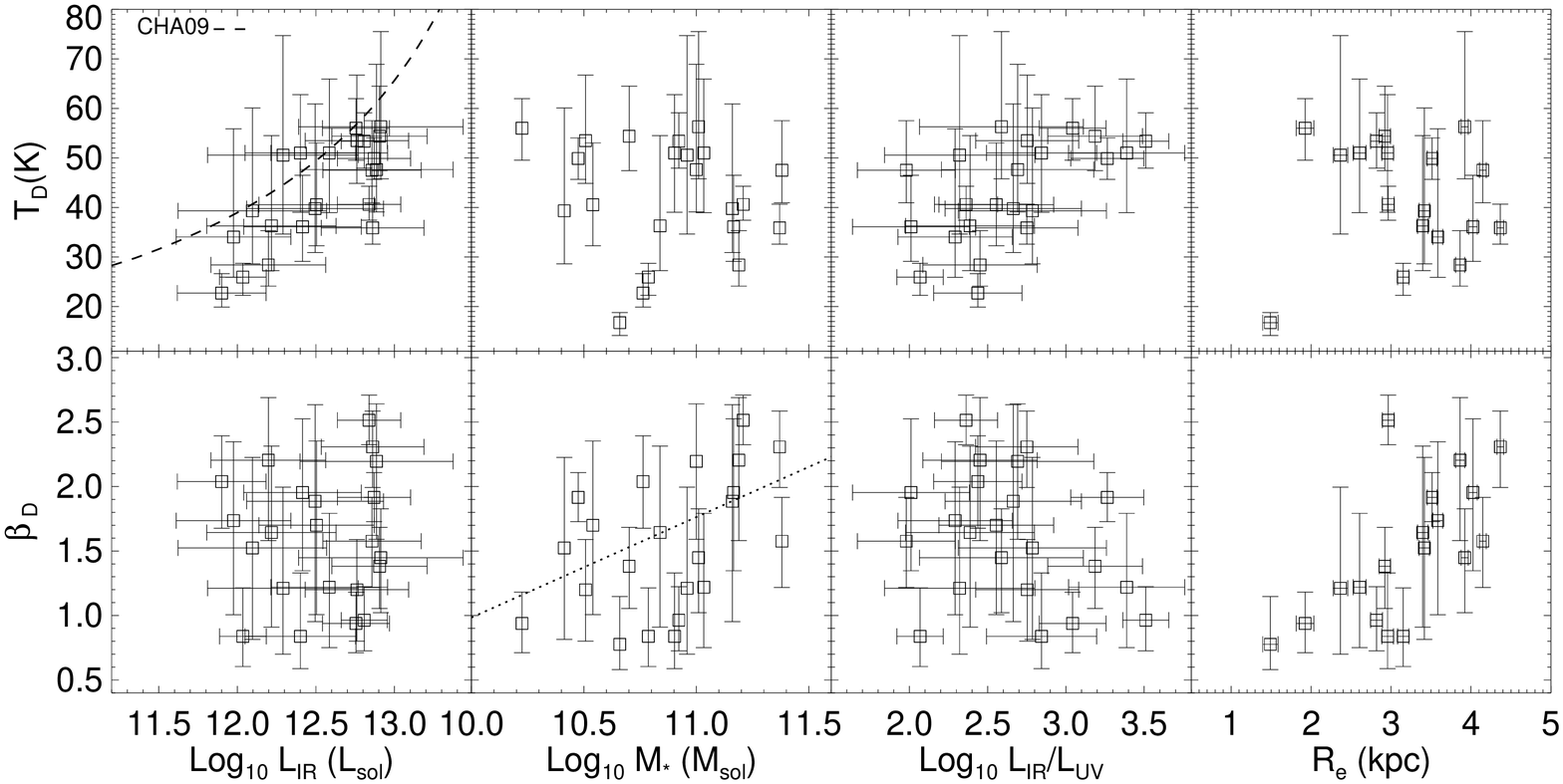}
\caption{Relationships between submm SED fit parameters and galaxy physical properties. Only those 450$\,\mu$m sources with reliable 850$\,\mu$m and SPIRE photometry are considered, leaving a total of 24 submm sources. Correlations between several of the parameters are seen. The best-fit relation for the $L_{\rm IR}-T_{\rm D}$ relation for {\it IRAS} galaxies from Chapin, Hughes \& Aretxaga (2009; CHA09) is shown in the top left panel. The dotted line in the bottom-middle panel is the best fit linear relation found between $\log_{10}M_{\ast}$ and $\beta_{\rm D}$.}
\label{fig:sedpars}
\end{figure*}

Nonetheless, we consider the possibility that this spread in the SED shape parameters is in part a function of underlying galaxy properties (as opposed to just random scatter). Fig.~\ref{fig:sedpars} shows the relationship between the submm SED parameters ($T_{\rm D}$ and $\beta_{\rm D}$) and the galaxy properties: IR luminosity ($L_{\rm IR}$, integrated in the range 8--1000$\,\mu$m)), stellar mass ($M_{\ast}$), dust attenuation (as probed by $L_{\rm IR}/L_{\rm UV}$ and effective radius ($R_{\rm e}$). Stellar mass estimates are made by finding the best fit to the optical/near-IR photometry from a grid of Bruzual \& Charlot (2003) templates (solar metallicity, Chabrier 2003 IMF) assuming the photometric redshift as described in Cirasuolo et al.\ (2010) and Micha{\l}owski et al.\ (2012). The UV luminosity is estimated via $\lambda L_{\lambda}$ from a power-law fit to the observed rest-frame UV photometry, with $\lambda=160\,$nm. The effective radius is calculated using the measured $R_{50}$ (i.e. the radius which contains 50 per cent of the total light) in the {\it HST} $F160W$ imaging, and the angular diameter distance at the assumed photometric redshift. 

Weak correlations between the submm SED shape parameters and some of the galaxy properties can be seen. In order to determine the statistical significance of these correlations, Spearman rank correlation coefficients are calculated for each of the combinations of submm SED and galaxy properties shown in Fig.~\ref{fig:sedpars}. These correlation coefficients, and their statistical deviation from the null hypothesis of no correlation, are tabulated in Table \ref{tab:srseds}. Before interpreting these correlations it is again worth noting that the co-variance between $T_{\rm D}$ and $\beta_{\rm D}$ means that we cannot be sure which of these parameters is driving the correlations with other observables, only that the mean submm SED shape is weakly dependant on them.

The IR luminosity, $L_{\rm IR}$, is seen to correlate with $T_{\rm D}$. The existence of the $L_{\rm IR}-T_{\rm D}$ relation has been well-documented since {\it IRAS} (Soifer \& Neugebauer 1991; Chapman et al.\ 2003; Chapin, Hughes \& Aretxaga 2009). Recent {\it Herschel} results show that this relationship extends to high redshift (i.e. $z\sim2$; Hwang et al.\ 2010; Magdis et al.\ 2010; Symeonidis et al.\ 2013). We compare our estimate of the $L_{\rm IR}-T_{\rm D}$ relation to that determined for local {\it IRAS} galaxies by Chapin, Hughes \& Aretxaga (2009; henceforth CHA09), where we have converted between {\it IRAS} colour and $T_{\rm D}$ using Eq.~\ref{eqn:mbb}, and fixed values of $c/\nu_0=100\,\mu$m and $\beta_{\rm D}=1.6$. Reasonable agreement is seen between our $L_{\rm IR}-T_{\rm D}$ relation and the CHA09 one, although the 450$\,\mu$m sources are slightly biased to colder values than the CHA09 relation. This could either be a result of true evolution towards colder dust at high-$z$ (e.g. Seymour et al.\ 2010), or a bias introduced by requirement of good submm photometry.

\begin{table}
\caption{Spearman rank correlation coefficients for the relationships between submm SED shape parameters ($T_{\rm D}$ and $\beta_{\rm D}$) and the galaxy properties, $L_{\rm IR}$, $M_{\ast}$ and $R_{\rm e}$. The probability of achieving these correlation coefficients by chance (i.e. the probability of the observed data if no correlation exists) is given in parentheses.}
\label{tab:srseds}
\begin{tabular}{lrr}
\hline
 & $T_{\rm D}$ & $\beta_{\rm D}$\\
\hline\hline
$L_{\rm IR}$ & \  \  0.71 (0.001) &\ 0.30 (0.40)\\
$M_{\ast}$ & $-0.30$\ \ \ \  (0.2) & \ 0.49 (0.02) \\
$L_{\rm IR}/L_{\rm UV}$&0.4\ \ \ \ (0.06) &-0.40\ (0.06)\\
$R_{\rm e}$ & $-0.16$\  \  \  \ (0.5)&\ 0.41 (0.05) \\
\hline
\end{tabular}
\end{table}


Moving onto the stellar mass, a tentative correlation is seen between $M_{\ast}$ and $\beta_{\rm D}$. Putting aside the physical implications of such a relation for now, the existence of an $M_{\ast}-\beta_{\rm D}$ relation allows us to predict $\beta_{\rm D}$ from a parameter unassociated with the IR SED. We fit a linear relation to $\log_{10}M_{\ast}$-$\beta_{\rm D}$, finding a best fit relation of $\beta_{\rm D}=(-7\pm2)+(0.8\pm0.2)\log_{10}M_{\ast}$. This relation will be used later to fix $\beta_{\rm D}$ in the SED fits of SCUBA-2 sources that lack reliable submm photometry.


Comparing the dust temperature and emissivity to the dust attenuation, as probed by the ratio of IR-to-UV luminosity reveals some weak correlations, with $T_{\rm D}$ appearing to vary with increasing $L_{\rm IR}/L_{\rm UV}$, while $\beta_{\rm D}$ anti-correlates with attenuation. Both of these correlations are statistically weak, with a Spearman rank $p=0.06$. Finally, a correlation between effective radius and $\beta_{\rm D}$ is seen, with $R_{\rm e}$ seen to increase with increasing $\beta_{\rm D}$.  

With this information in hand, we now perform SED fits to the remaining 31 SCUBA-2 450$\,\mu$m sources with reliable identifications and redshifts. The main purpose of this fitting is not to determine parameter values, but merely to have a template with which to estimate the integrated IR luminosity. Again we fit a modified blackbody of the form seen in Eq.~\ref{eqn:mbb}, allowing only $L_{\rm IR}$ and $T_{\rm D}$ to vary, with fixed values for the other parameters: $\nu_0=c/100\,\mu$m and $\beta_{\rm D}=-7+0.8\log_{10}M_{\ast}$. The IR luminosities ($L_{\rm IR}$) estimated from this fitting, along with all the other galaxy properties, are given in Table \ref{tab:s2details}.

\subsection{The potential for a stellar mass -- dust emissivity relation }\label{sec:smbeta}

The results shown in Fig.~\ref{fig:sedpars} suggest that the shape of the submm SED is somewhat sensitive to the gross properties of the galaxy, with tentative correlations seen between $\beta_{\rm D}$ and both stellar mass and size. Interpretation of these correlations is difficult for a number of reasons. Firstly, a single modified blackbody SED is clearly not a complete physical model for the dust emission in a galaxy; a typical star forming galaxy will contain many distinct star-forming regions enshrouded by dust, as well as more diffuse dust heated by evolved stars, and our observed submm SED is the integrated emission from all of these components. Secondly, the parameters $T_{\rm D}$ and $\beta_{\rm D}$ are known to be degenerate in modified blackbody fits to submm data limited in both wavelength range and signal-to-noise (Shetty et al.\ 2009a). Indeed, the 2D posteriors recovered by our MCMC fits for the 23 sources shown in Fig.~\ref{fig:sedpars} typically show a correlation coefficient between $T_{\rm D}$ and $\beta_{\rm D}$ of $\sim-0.9$. 

However, the existence of this degeneracy between $T_{\rm D}$ and $\beta_{\rm D}$ does not invalidate the conclusion that there may be a relationship between the submm SED shape and both stellar mass and radius. To confirm this, we perform a simple Monte-Carlo simulation, taking the measured stellar masses and radii for the 23 sources in Fig.~\ref{fig:sedpars} and generating random realisations of $T_{\rm D}$ and $\beta_{\rm D}$ which are imprinted with the covariance found in the MCMC fits to the real data. We produce 10,000 realisations of these simulations and search for correlations of similar strength to those seen in the real data. As expected, the incidence of correlations at the strength quoted in Table \ref{tab:srseds} is close to what would be predicted analytically; $p=0.017\pm0.001$ for the $M_{\ast}-\beta_{\rm D}$ relation, and $p=0.044\pm0.001$ for the $R_{\rm e}-\beta_{\rm D}$ relation.


Thus while the correlations between SED shape and both stellar mass and effective radii may be real, to observationally discriminate which of the SED parameters is driving this relationship would require the individual star forming regions to be resolved at a wide range of far-IR and submm wavelengths, something that not even ALMA will be capable of for these distant galaxies. However, we can check if this behaviour is also seen in resolved studies of nearby galaxies with {\it Herschel}. Galametz et al.\ (2012) present a detailed analysis of the dust temperature and emissivity distribution observed within eleven local galaxies using data from the {\it Herschel} KINGFISH survey (Kennicutt et al.\ 2011). By fitting single modified blackbody SEDs to individual pixels in the {\it Herschel} images they find a wide range of dust temperatures and emissivities within galaxies. In many cases a trend of decreasing dust temperature and emissivity is seen with radius. These trends are also seen in {\it Herschel} observations of M31 (Smith et al.\ 2012b). Again these studies are subject to the same caveats about $T_{\rm D}-\beta_{\rm D}$ degeneracies, although the resolved nature of the observations, as well as the increased signal-to-noise, alleviates some of these concerns.

\begin{figure}
\includegraphics[scale=0.45]{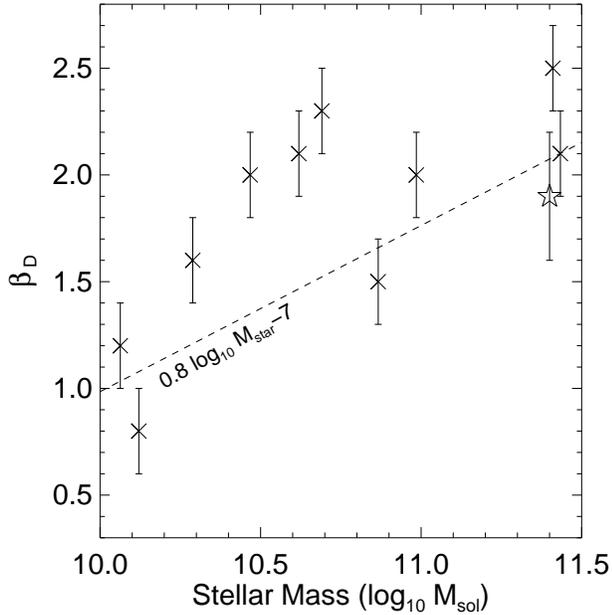}
\caption{Dust emissivity vs. stellar mass for local galaxies in the KINGFISH sample (crosses; Galametz et al.\ 2012) and M31 (star; Smith et al.\ 2012). Overplotted is the $M_{\ast}-\beta_{\rm D}$ relation seen in our SCUBA-2 sample in Fig.~\ref{fig:sedpars}.}
\label{fig:lbm}
\end{figure}

If we take the median emissivity values for the KINGFISH sample (Table 3 of Galametz et al.\ 2012) and M31 (Smith et al.\ 2012b) and compare these to the stellar masses for these local galaxies (estimated from the published $K_{\rm s}$ magnitudes and assuming $M_{\ast}/L_{K_{\rm s}}=0.8$; Munoz-Mateos et al.\ 2009; Bell et al.\ 2003) we again see a trend of dust emissivity increasing with stellar mass. This is shown in Fig.~\ref{fig:lbm}. The emissivity--stellar mass relation for local, resolved, galaxies is nearly identical to the one we see for distant 450$\,\mu$m-selected sources. This should not be taken as evidence for a real physical link between these parameters, merely that the submm SED shape for both local and distant (i.e. $z>1$) galaxies responds in a similar way to changes in the stellar mass.


\subsection{The specific star formation rate of 450$\,\mu$m sources}\label{sec:ssfr}

The primary utility of deep submm imaging is as an unambiguous tracer of SFR. The ability to probe rest-frame wavelengths longward of 100$\,\mu$m is particularly valuable for distant galaxies as it is difficult for physical processes other than star formation (i.e. AGN, diffuse ISM dust) to be responsible for the highly luminous, but still cold, thermal emission observed (e.g. Fig.~\ref{fig:seds}). This allows us to convert our integrated IR luminosities into SFRs without reservation, using the relation of Kennicutt et al.\ (1998), after converting to a Chabrier (2003) IMF; $\log_{10}$(SFR/M$_{\odot}\,$yr$^{-1})=\log_{10}(L_{\rm IR}/{\rm L}_{\odot})-10$.

\begin{figure}
\includegraphics[scale=0.37]{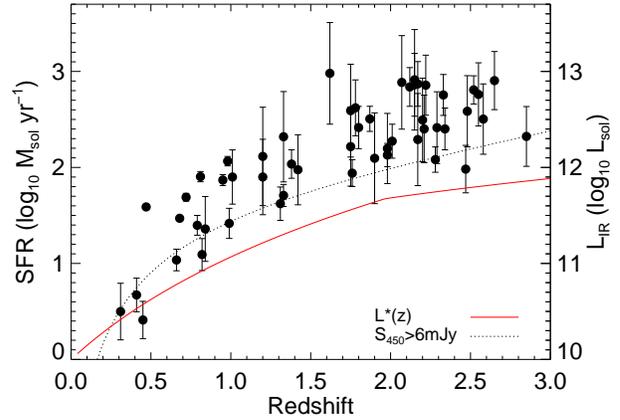}
\caption{SFR vs. redshift for SCUBA-2 450$\,\mu$m sources. The dotted lines is the expected minimum SFR (LIR) as a function of redshift, assuming a modified blackbody SED with the mean parameters from \S\ref{sec:seds} (i.e. $T_{\rm D}=42$\,K, $\beta_{\rm D}=1.6$). The solid red line is $L^{*}(z)$ as estimated by Gruppioni et al.\ (2013). }
\label{fig:lirz}
\end{figure}

In Fig.~\ref{fig:lirz} we plot the SFR estimates for our SCUBA-2 450$\,\mu$m sources against redshift. Our SCUBA-2 data set is clearly limited to SFRs of greater than 10\,M$_{\odot}$\,yr$^{-1}$ at $z=1$ and 100\,M$_{\odot}$\,yr$^{-1}$ at $z=2$. It is interesting to compare these limits to the evolution of the knee of the IR luminosity function, $L^{*}$. Also shown in Fig.~\ref{fig:lirz} is the evolution of $L^*$ with redshift as estimated by Gruppioni et al.\ (2013). Because of the combined effect of the negative $K$-correction and strong evolution in the IR luminosity function the SCUBA-2 450$\,\mu$m sensitivity limit is only slightly, although consistently, above $L^{*}(z)$ with a difference of similar to 0.3\,dex, i.e. a factor of roughly two for $z>0.5$.

Thus with our current SCUBA-2 dataset we are not able to individually detect "typical" galaxies at high redshift. However the small separation between our sensitivity limits and $L^*(z)$, and the quality of the ancillary CANDELS data in our S2CLS fields, means that the aggregate SFRs of galaxies at all redshifts should be easily obtainable via stacking techniques.

In both the UDS and COSMOS fields we take the full $F160W<25$ AB mag CANDELS catalogues with photometric redshifts and stellar mass estimates. These catalogues are split by both redshift and mass into six bins described in Table \ref{tab:stack}. Passive galaxies are excluded by requiring that the rest-frame NUV$-r<3.5$, as per Karim et al.\ (2011).

Table \ref{tab:stack} gives the numbers of sources and mean quantities for these six bins. Stacking is performed in the four-submm bands available, using the method described in Appendix \ref{sec:stacking}. Estimates on the variance in the stacked quantities are calculated by jack knifing the input catalogue.

\begin{table}
\caption{Details of $z$--$M_{\ast}$ bins used to stack the submm imaging.}
\label{tab:stack}
\begin{tabular}{lrrr}
\hline
$z$ & $M_{\ast}$ & $N_{\rm src}$ & $\langle L_{\rm IR} \rangle$\\
 & $\log_{10}M_{\odot}$ & & $\log_{10}L_{\odot}$ \\
\hline\hline
$0<z<1.5$ & $<10$ $\langle9.\rangle$& 4466 & $9.8\pm0.2$ \\
$0<z<1.5$ & $10...10.5$ $\langle10.2\rangle$& 314 & $11.3\pm0.1$\\
$0<z<1.5$ & $>10.5$$\langle10.7\rangle$ & 163 & $11.5\pm0.2$\\
$1.5<z<3$ & $<10$ $\langle9.7\rangle$& 2270 & $9.6\pm1.0$\\
$1.5<z<3$ & $10...10.5$ $\langle10.2\rangle$ & 527 & $11.6\pm0.2$\\
$1.5<z<3$ & $>10.5$ $\langle10.7\rangle$ & 406 & $11.8\pm0.2$\\
\hline
\end{tabular}
\end{table}

\begin{figure}
\includegraphics[scale=0.35,clip=0cm 2cm 0cm 0cm]{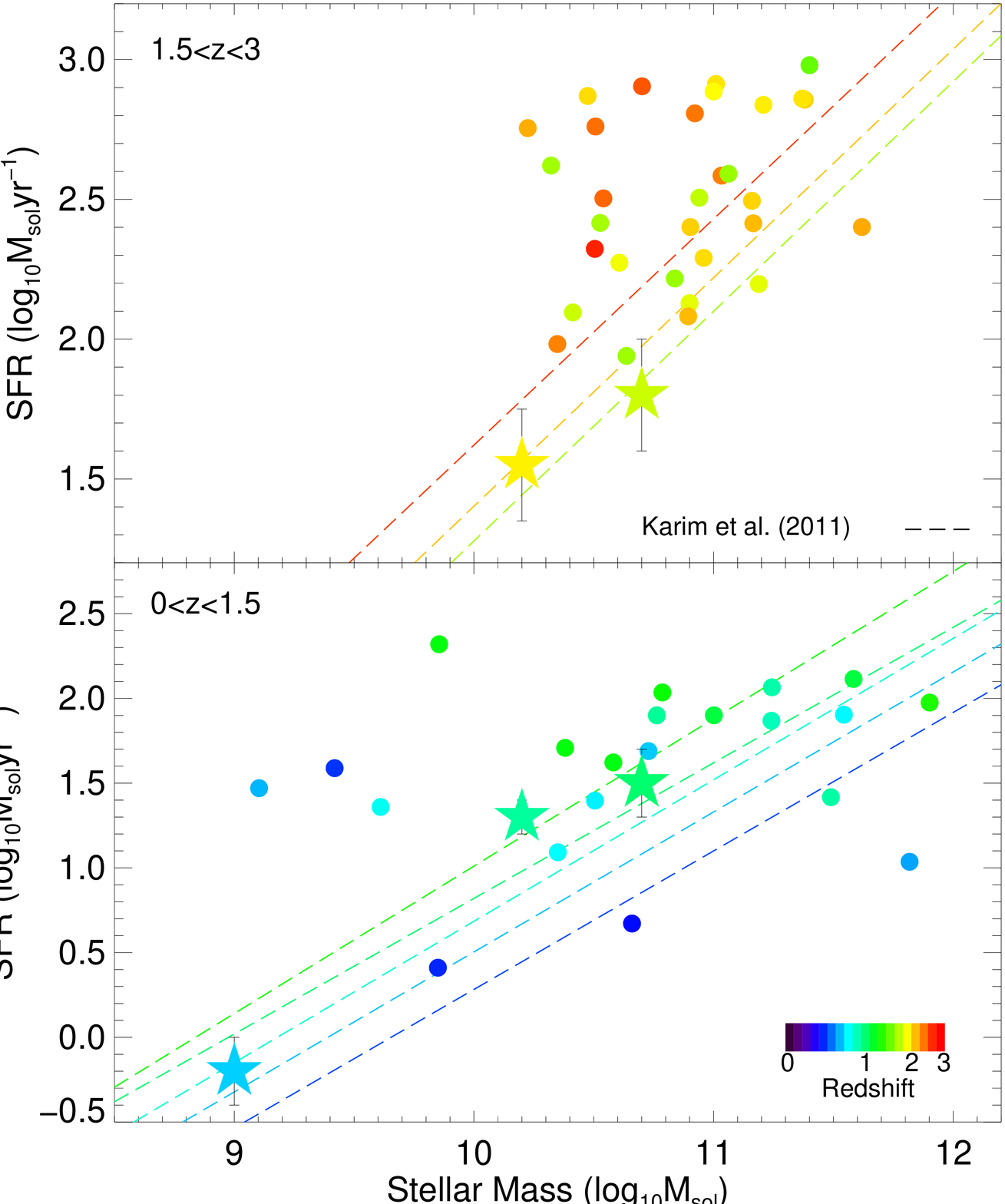}
\caption{SFR vs. stellar mass for 450$\,\mu$m sources. Sources are split into two redshift bins ($0<z<1.5$ and $1.5<z<3$) and colour-coded by redshift. Individual sources detected at 450$\,\mu$m are shown as filled circles, while results from stacking the submm data of near-IR selected galaxies in mass and redshift bins are shown as filled stars. Shown for comparison is the mean SFR-$M_{\ast}$ relation for star-forming galaxies as a function of redshift from Karim et al.\ (2011). }
\label{fig:sfrsm}
\end{figure}

Figure \ref{fig:sfrsm} compares the SFR to the stellar mass for both individual galaxies in our 450$\,\mu$m sample, and our stacked quantities. Shown for comparison in Fig.~\ref{fig:sfrsm} are the mean SFR--$M_{\ast}$ relations seen for star-forming galaxies at a variety of redshifts via stacking in the radio (Karim et al.\ 2011).  In general our stacked results show good agreement with the Karim et al.\ (2011) SFR-$M_{\ast}$ relation, or typical specific SFR (sSFR: SFR/$M_{\ast}$) at a given redshift. Meanwhile the individually detected 450$\,\mu$m sources tend to lie above this, a result of the limited sensitivity of, and the small volume probed by, our sample.

In recent years the dispersion around the SFR-$M_{\ast}$ relation has become a quantity of much interest, as the existence of a distinct population of ``starbursts'' with exceptionally high sSFR has been postulated. Using parametric fits to the spread in the SFR--$M_{\ast}$, Rodighiero et al.\ (2011) suggest that the ``main sequence'' population (i.e. those galaxies consistent with the SFR--$M_{\ast}$ relation) can be described by a Gaussian distribution with $\sigma=0.25\,$dex plus a population of  excess ``starbursts''  observed at +0.6\,dex.

Using the same definition for starbursts here, we identify those individual 450$\,\mu$m sources which have a sSFR +0.6\,dex, or more, greater than the stacked sSFR  at comparable redshift and mass . Of the 54 sources considered, only 14 are classified as starbursts (26 per cent). This fraction is reasonably consistent across the two redshift bins, with 5/23 (22 per cent) starbursts in the low-$z$ bin and 9/31 starbursts (29 per cent) in the high-$z$ bin. Rodighiero et al.\ (2011) estimate that the starburst fraction is $20-30$ per cent for samples limited to SFR$\gs100\,{\rm M}_{\odot}\,$yr$^{-1}$, in good agreement with our estimate.

However, this definition for a population of excess sSFR starbursts is somewhat arbitrary, and highly sensitive to the assumed location, and width, of the SFR-$M_{\ast}$ relation. While our data are not substantial enough to perform parametric fits to the full SFR-$M_{\ast}$ space, we can ask whether the sSFR values determined for our individual sources, in combination with our stacking results, are consistent with a single uni-modal distribution. 

For each 450$\,\mu$m source we estimate the mean sSFR at its redshift and mass by extrapolating from our stacked sSFR values, and assuming the mean slope of the SFR-$M_{\ast}$ found by Karim et al.\ (2011), i.e. SFR$\propto M_{\ast}^{0.4}$. As our sample covers a wide range in redshift, across which the mean SFR-$M_{\ast}$ relation and our sensitivity to $L_{\rm IR}$ is varying it is difficult to correct our sample for completeness in terms of distance from the SFR-$M_{\ast}$ relation, i.e. $\Delta$sSFR. To avoid issues of incompleteness in terms of the stellar mass we consider only the 47 sources with $M_{\ast}>10^{10}\,$M$_{\odot}$.  However, we can crudely estimate a sSFR at which our sample becomes complete using the results of Figures ~\ref{fig:lirz} and \ref{fig:sfrsm}. In the redshift range $0<z<1.5$ our limiting SFR is $>40$\,M$_{\odot}$\,yr$^{-1}$, while at  $1.5<z<3$ it is $>160$\,M$_{\odot}$\,yr$^{-1}$. Given our stacking results the mean sSFR for $M_{\ast}=10^{10.2}\,$M$_{\odot}$ galaxies is 0.1 and 0.4\,Gyr$^{-1}$ for the low and high-$z$ bins, respectively. Taking our estimated SFR limits and the mean stellar mass in these bins (10$^{10.2}\,$M$_{\odot}$) our limiting sSFR is $\sim0.4$ and $\sim1$\,Gyr$^{-1}$. So we can expect that we are roughly complete in terms of $\Delta$sSFR at $>0.3$ in the low-$z$ bin and $>0.6$ in the high-$z$ bin, i.e. below the threshold of $+0.6$\,dex for starbursts set by Rodighiero et al.\ (2011).

For galaxies with $M_{\ast}>10^{10}\,$M$_{\odot}$ we observe nine with a sSFR excess of +0.6\,dex or more. A Gaussian distribution with $\sigma=0.25$\,dex normalised to the total number of  $M_{\ast}>10^{10}\,$M$_{\odot}$ predicts that we should observe six, so our result represents a $\sim1\sigma$ excess assuming Poisson statistics. While we are clearly limited by our small sample, we do not find any evidence that these high sSFR galaxies belong to a distinct population from the general population of star-forming galaxies. It is worth noting that at the mass limit considered here (i.e. $M_{\ast}>10^{10.5}\,$M$_{\odot}$), Rodighiero et al.\ (2011) find significantly smaller numbers of high sSFR galaxies than at lower masses. It may be that the starburst phenomena is more prevalent at masses lower than we can reasonably probe here.

\begin{figure}
\includegraphics[scale=0.45]{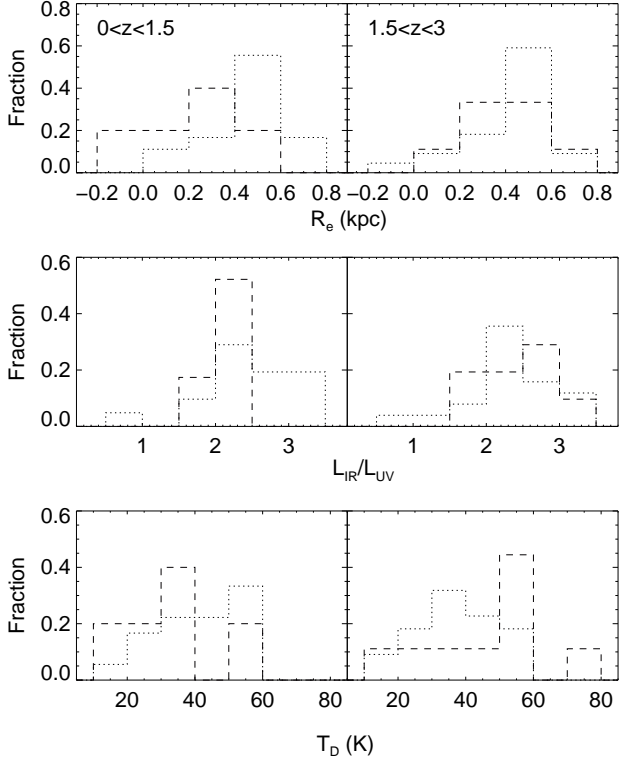}
\caption{Distribution of galaxy properties for star-burst (dashed) and normal star-forming (dotted) galaxies.}
\label{fig:sbprops}
\end{figure}

Another way to test the sSFR--starburst hypothesis is to look for correlations between sSFR and other galaxy properties. In Fig.~\ref{fig:sbprops} we compare the galaxy properties considered here that are unrelated to specific SFR (i.e. $R_{\rm e}$, $L_{\rm IR}/L_{\rm UV}$ and $T_{\rm D}$) for ``starburst'' (i.e. $\Delta$sSFR$>0.6$\,dex) and normal star-forming galaxies. While there are some hints of differences, with starbursts potentially hotter and experiencing more dust attenuation (i.e. higher $L_{\rm IR}/L_{\rm UV}$), these differences are not statistically significant; KS tests suggest that all of these distributions are consistent at the $p=0.5$ level or better. While this may be a result of the small sample size considered here, some larger studies also fail to find significant trends between location in the SFR--$M_{\ast}$ plane and galaxy properties. Law et al.\ (2012) considered 306 distant galaxies with {\it HST} WFC3 and found no morphological dependence on specific SFR. Taking samples of close pairs from $0<z<1$, Xu et al.\ (2012) found no evidence for an increase in specific SFR for interactions vs. normal isolated disk galaxies at $z\sim1$, with a preference for interactions to have high specific SFR only seen at low-$z$. Near-IR IFU studies also fail to find a connection between internal kinematics (i.e. rotation or dispersion dominated) and being on or off the ``main sequence'' (F\"{o}rster-Schreiber et al.\ 2009).

This connection (or lack of) between specific SFR and galaxy properties is important for a number of reasons. The dominance of the SFR--$M_{\ast}$ relation has been taken as evidence that the influence of mergers in galaxy formation is minor (Rodighiero et al.\ 2011). However, there is no clear observational evidence that galaxy-galaxy mergers should lie above the SFR--$M_{\ast}$ relation. Modern hydrodynamical simulations suggest that while mergers can boost SFR, these enhancements may only be modest in some cases (i.e. factors of two) and occur over timescales significantly shorter than the observational merger signatures (di Matteo et al.\ 2008; Hopkins et al.\ 2011; Bournaud et al.\ 2011; Hopkins et al.\ 2012; Hayward et al.\ 2012). For these reasons, using sSFR as a tracer for the number of merger-induced starbursts is flawed; while high values of sSFR may be exclusively merger-driven, not all observed mergers will have high sSFR. This distinction is also consistent with the discrepancy between the fraction of high sSFR starbursts ($\sim2$ per cent), and the global merger rate at $z\sim2$ ($10$--$20$ per cent; Conselice et al.\ 2000; Lotz et al.\ 2011; Law et al.\ 2012; Bluck et al.\ 2012). Future resolved observations of distant galaxies with facilities like ALMA will likely determine if the sSFR of galaxies is indeed a signature of merger-induced star-formation.

\section{Conclusions}\label{sec:conc}
We have presented a detailed investigation into the multi-wavelength properties of 69 450-$\mu$m sources selected from two fields (COSMOS \& UDS) from the SCUBA-2 Cosmology Legacy Survey (S2CLS). These data represent the first 10 months of observing for S2CLS and reach a depth of $\sim1.5$\,mJy RMS at 450$\,\mu$m in the deepest regions. Using likelihood ratio techniques we found robust optical/near-IR counterparts from overlapping {\it HST} CANDELS data for 58 (84 per cent) of these sources. Focusing on this sample of robustly identified sources we were able to reach the following conclusions;
\begin{itemize}
\item The redshift distribution of 450$\,\mu$m sources with $S_{450}\gs6\,$mJy was estimated, with a broad peak in $dN/dz$ seen between $1<z<3$ and an overall median of $z=1.4$. Few 450$\,\mu$m sources are found to lie at very high redshifts (i.e. $z>3$). While the 26 per cent of sources without identifications could lie at preferentially higher redshifts, their submm flux densities and colours suggest that they are not distinct from the sources with identifications.
\item By fitting modified blackbody SEDs to the combined {\it Herschel} SPIRE and SCUBA-2 photometry, the range and physical dependance of SED properties on galaxy properties was investigated. Using an MCMC fitting technique, and allowing both the dust temperature ($T_{\rm D}$) and emissivity ($\beta_{\rm D}$) to vary, we found the mean values for these parameters for our 450$\,\mu$m sample ($\langle L_{\rm IR}\rangle= 10^{12}\,{\rm L}_{\odot}$) to be $\langle T_{\rm D}\rangle=42\pm11$\,K and $\langle \beta_{\rm D}\rangle=1.6\pm0.5$. Comparing these SED parameters; clear evidence of the well-known $L_{\rm IR}$--$T_{\rm D}$ relation was observed, while new weak correlations between $\beta_{\rm D}$ and both stellar mass and effective radius were discovered. These relations all appear to be similar to those observed for local galaxies.
\item The relationship between star formation rate and stellar mass was investigated, with the majority (74 per cent) of 450$\,\mu$m sources found to be consistent with previous measures of the SFR--$M_{\ast}$ relation (e.g. Karim et al.\ 2011). Apart from their specific SFR, these ``starburst'' galaxies are indistinguishable from the galaxies on the SFR--$M_{\ast}$ relation, although this may be a result of the limited diagnostic power of the multi-wavelength data we possess for these sources.

\end{itemize} 

When completed, S2CLS will possess deep ($\sigma_{450}\sim1\,$mJy) 450$\,\mu$m imaging over $0.6$ deg.$^2$; a factor of almost ten over the area used in this work. All of these data will have overlapping multi-wavelength coverage similar to that available for the fields studied here. On the basis of our results, the potential for 450$\,\mu$m imaging from S2CLS is clear; a large sample ($\sim 600$) of submm selected galaxies with exquisite SED information from the optical through to submm will allow detailed questions about the nature of dust and the role of starbursts in distant galaxies to be addressed in a way not previously possible, and likely to be unsurpassed until the arrival of large-scale ALMA surveys and new single-dish facilities like LMT and CCAT (Hughes et al.\ 2010; Woody et al.\ 2012). 

\section*{Acknowledgements}
We thank the anonymous referee for suggestions which greatly enhanced this work.\\


We thank B. Sidthorpe and W. Holland for many useful discussions related to the reduction and analysis of SCUBA-2 imaging.\\
JSD acknowledges the support of the Royal Society via a Wolfson Research Merit award, and the support of the European Research Council via the award of an Advanced Grant.\\
IRS acknowledges support from STFC, a Leverhulme Fellowship, the ERC Advanced Investigator programme DUSTYGAL and a Royal Society/Wolfson Merit Award.\\
LW and SJO were supported by the Science and Technology Facilities Council [grant number ST/I000976/1].\\
Many thanks the JCMT telescope operators; J. Hoge, J. Wouterloot and W. Montgomerie, without whom these observations would
not be possible. \\

The James Clerk Maxwell Telescope is operated by the Joint Astronomy Centre on behalf of the Science and Technology Facilities Council of the United Kingdom, the National Research Council of Canada, and (until 31 March 2013) the Netherlands Organisation for Scientific Research. Additional funds for the construction of SCUBA-2 were provided by the Canada Foundation for Innovation.\\

SPIRE has been developed by a consortium of institutes led
by Cardiff Univ. (UK) and including: Univ. Lethbridge (Canada);
NAOC (China); CEA, LAM (France); IFSI, Univ. Padua (Italy);
IAC (Spain); Stockholm Observatory (Sweden); Imperial College
London, RAL, UCL-MSSL, UKATC, Univ. Sussex (UK); and Caltech,
JPL, NHSC, Univ. Colorado (USA). This development has been
supported by national funding agencies: CSA (Canada); NAOC
(China); CEA, CNES, CNRS (France); ASI (Italy); MCINN (Spain);
SNSB (Sweden); STFC, UKSA (UK); and NASA (USA).\\

\appendix
\section{SCUBA-2 450$\,\mu$\lowercase{m} source catalogues}
\onecolumn
\footnotesize
\begin{longtable}{lllllllllllll}
\caption{450$\,\mu$m sources in COSMOS CANDELS region. Sources with only unreliable identifications (i.e. $\log_{10}LR<1.1$) are shown in italics. Where multiple good identifications for sources exist these are listed with the suffix .2, .3, etc.}\\
\label{tab:s2cosmos}
 ID & RA$_{450}$ & Dec$_{450}$ & $S_{450}$ & $\Delta S_{450}$ & SNR$_{450}$ &  $S_{850}$ & $\Delta S_{850}$ & RA$_{\rm opt}$ & Dec$_{\rm opt}$ & Sep. & $\log_{10}LR$ & $z_{\rm phot}$\\
 \hline
  & deg. & deg. & mJy & mJy & & mJy & mJy & deg. & deg. & arcsec & & \\
\hline\hline
\endfirsthead
\caption{(continued)}\\
 ID & RA$_{450}$ & Dec$_{450}$ & $S_{450}$ & $\Delta S_{450}$ & SNR$_{450}$ &  $S_{850}$ & $\Delta S_{850}$ & RA$_{\rm opt}$ & Dec$_{\rm opt}$ & Sep. & $\log_{10}LR$ & $z_{\rm phot}$\\
 \hline
  & deg. & deg. & mJy & mJy & & mJy & mJy & deg. & deg. & arcsec & & \\
\hline\hline
\endhead
\hline
\endfoot
S2CLSc01 & 150.1638 &  2.3721 & 24.46 &  1.77 & 13.72 &  3.19 &  0.34 & 150.16357 &   2.37242 &  1.38 &  4.16 &  2.12\\
S2CLSc02 & 150.1055 &  2.3125 & 18.91 &  1.42 & 13.13 &  5.18 &  0.21 & 150.10546 &   2.31285 &  1.09 &  4.12 &  2.65\\
S2CLSc03 & 150.1430 &  2.3556 & 15.17 &  1.40 & 10.55 &  4.67 &  0.23 & 150.14304 &   2.35585 &  0.89 &  3.90 &  2.33\\
S2CLSc04 & 150.0985 &  2.3207 & 14.67 &  1.43 & 10.06 &  4.22 &  0.22 & 150.09866 &   2.32081 &  0.87 &  3.91 &  2.55\\
S2CLSc04.2 & 150.0985 &  2.3207 & 14.67 &  1.43 & 10.06 &  4.22 &  0.22 & 150.09914 &   2.32069 &  2.46 &  2.12 &  1.03\\
S2CLSc05 & 150.0989 &  2.3649 & 16.74 &  1.67 &  9.83 &  5.57 &  0.24 & 150.09854 &   2.36536 &  1.99 &  3.53 &  2.52\\
S2CLSc06 & 150.0655 &  2.2636 & 20.91 &  2.27 &  8.98 & 10.41 &  0.32 & 150.06460 &   2.26405 &  3.59 &  1.17 &  1.38\\
S2CLSc07 & 150.1000 &  2.2971 & 14.73 &  1.64 &  8.79 &  6.87 &  0.24 & 150.10014 &   2.29713 &  0.55 &  3.47 &  0.41\\
S2CLSc08 & 150.1065 &  2.2516 & 18.11 &  2.24 &  8.00 &  3.48 &  0.31 & 150.10641 &   2.25161 &  0.37 &  4.62 &  2.17\\
S2CLSc08.2 & 150.1065 &  2.2516 & 18.11 &  2.24 &  8.00 &  3.48 &  0.31 & 150.10635 &   2.25110 &  1.82 &  2.38 &  0.66\\
S2CLSc08.3 & 150.1065 &  2.2516 & 18.11 &  2.24 &  8.00 &  3.48 &  0.31 & 150.10587 &   2.25185 &  2.48 &  2.00 &  1.66\\
S2CLSc09 & 150.0568 &  2.3730 & 18.21 &  2.34 &  7.71 &  3.38 &  0.31 & 150.05657 &   2.37375 &  2.72 &  3.01 &  1.01\\
S2CLSc10 & 150.1504 &  2.3635 & 11.98 &  1.60 &  7.42 &  2.80 &  0.28 & 150.15026 &   2.36414 &  2.30 &  3.14 &  2.29\\
S2CLSc11 & 150.1874 &  2.3226 & 11.97 &  1.65 &  7.18 &  2.16 &  0.29 & 150.18763 &   2.32250 &  0.95 &  3.90 &  1.42\\
S2CLSc12 & 150.1541 &  2.3276 & 10.09 &  1.39 &  7.18 &  2.89 &  0.23 & 150.15374 &   2.32800 &  1.93 &  2.91 &  2.48\\
S2CLSc13 & 150.0780 &  2.2819 & 13.35 &  1.91 &  6.94 &  3.19 &  0.27 & 150.07787 &   2.28116 &  2.60 &  3.00 &  1.98\\
S2CLSc14 & 150.1390 &  2.4327 & 21.58 &  3.31 &  6.43 &  6.10 &  0.52 & 150.13872 &   2.43226 &  1.78 &  2.46 &    -- \\
S2CLSc14.2 & 150.1390 &  2.4327 & 21.58 &  3.31 &  6.43 &  6.10 &  0.52 & 150.13926 &   2.43202 &  2.46 &  2.35 &  1.02\\
S2CLSc15 & 150.1096 &  2.2938 & 10.55 &  1.61 &  6.42 &  2.23 &  0.23 & 150.10909 &   2.29433 &  2.43 &  3.26 &  1.75\\
S2CLSc16 & 150.1235 &  2.3608 &  8.64 &  1.37 &  6.23 &  2.02 &  0.21 & 150.12294 &   2.36096 &  2.17 &  3.49 &  1.90\\
S2CLSc17 & 150.1040 &  2.3456 &  8.31 &  1.31 &  6.10 &  0.25 &  0.20 & 150.10353 &   2.34611 &  2.45 &  1.42 &  0.95\\
S2CLSc18 & 150.1219 &  2.3407 &  7.45 &  1.21 &  6.02 &  1.80 &  0.19 & 150.12181 &   2.34131 &  2.10 &  2.46 &  2.17\\
S2CLSc19 & 150.1837 &  2.3856 & 10.71 &  1.83 &  5.73 &  1.33 &  0.34 & 150.18410 &   2.38640 &  3.16 &  2.25 &  1.87\\
S2CLSc20 & 150.0859 &  2.2971 &  9.88 &  1.82 &  5.44 &  1.07 &  0.26 & 150.08579 &   2.29765 &  2.12 &  3.28 &  0.98\\
S2CLSc21 & 150.1059 &  2.3256 &  7.13 &  1.29 &  5.42 &  1.22 &  0.20 & 150.10531 &   2.32590 &  2.35 &  2.39 &  0.84\\
S2CLSc22 & 150.0577 &  2.2929 & 11.58 &  2.09 &  5.38 &  2.86 &  0.28 & 150.05706 &   2.29286 &  2.39 &  3.06 &  1.33\\
S2CLSc23 & 150.0862 &  2.3807 & 11.38 &  2.15 &  5.11 &  1.82 &  0.31 & 150.08599 &   2.38083 &  0.93 &  2.46 &  0.31\\
S2CLSc23.2 & 150.0862 &  2.3807 & 11.38 &  2.15 &  5.11 &  1.82 &  0.31 & 150.08633 &   2.38091 &  0.97 &  2.28 &  1.46\\
S2CLSc23.3 & 150.0862 &  2.3807 & 11.38 &  2.15 &  5.11 &  1.82 &  0.31 & 150.08626 &   2.38142 &  2.67 &  1.39 &  2.34\\
S2CLSc24 & 150.0945 &  2.3350 &  7.29 &  1.41 &  5.08 &  0.44 &  0.21 & {\it 150.09502 }&   {\it 2.33529} &  {\it 2.28} & {\it -1.64} & {\it  0.05}\\
S2CLSc25 & 150.0815 &  2.3397 &  8.43 &  1.67 &  5.02 &  1.88 &  0.23 & 150.08105 &   2.34011 &  2.07 &  2.77 &  0.45\\
S2CLSc26 & 150.0855 &  2.2899 &  9.23 &  1.83 &  4.97 &  2.78 &  0.25 & 150.08488 &   2.28967 &  2.47 &  2.14 &  1.76\\
S2CLSc27 & 150.0769 &  2.3794 & 10.82 &  2.17 &  4.87 &  2.19 &  0.31 &{\it  150.07720} &  {\it 2.38038} &  {\it 3.71} &  {\it 0.54} & {\it  2.56}\\
S2CLSc28 & 150.1223 &  2.3998 &  9.72 &  1.94 &  4.85 & -0.19 &  0.29 & 150.12226 &   2.39977 &  0.25 &  3.16 &  0.47\\
S2CLSc28.2 & 150.1223 &  2.3998 &  9.72 &  1.94 &  4.85 & -0.19 &  0.29 & 150.12200 &   2.39963 &  1.24 &  2.48 &  0.62\\
S2CLSc29 & 150.1594 &  2.2971 &  8.67 &  1.75 &  4.84 &  1.33 &  0.30 & 150.15933 &   2.29680 &  1.21 &  3.44 &  1.98\\
S2CLSc29.2 & 150.1594 &  2.2971 &  8.67 &  1.75 &  4.84 &  1.33 &  0.30 & 150.15969 &   2.29727 &  1.11 &  2.64 &  0.43\\
S2CLSc30 & 150.1571 &  2.3584 &  8.25 &  1.68 &  4.83 &  0.05 &  0.29 & 150.15747 &   2.35803 &  1.96 &  2.69 &  2.47\\
S2CLSc31 & 150.1177 &  2.3297 &  5.93 &  1.22 &  4.82 &  1.62 &  0.19 & 150.11754 &   2.32996 &  1.00 &  3.81 &  2.28\\
S2CLSc32 & 150.1011 &  2.3344 &  6.48 &  1.31 &  4.82 &  4.34 &  0.20 & {\it 150.10119 }&  {\it  2.33437 }& {\it  0.48 }& {\it -0.56 }&    -- \\
S2CLSc33 & 150.0985 &  2.2601 &  9.73 &  2.00 &  4.77 &  3.94 &  0.28 & 150.09790 &   2.26001 &  2.20 &  1.75 &    -- \\
S2CLSc34 & 150.1349 &  2.3991 &  8.90 &  1.81 &  4.76 &  2.24 &  0.29 & 150.13513 &   2.39942 &  1.52 &  3.11 &  0.07\\
S2CLSc35 & 150.1126 &  2.3750 &  8.21 &  1.68 &  4.75 &  2.98 &  0.25 & 150.11240 &   2.37525 &  1.29 &  2.97 &  2.21\\
S2CLSc36 & 150.0869 &  2.3082 &  8.33 &  1.75 &  4.73 & -0.45 &  0.26 & 150.08701 &   2.30901 &  2.81 &  0.61 &  0.71\\
S2CLSc37 & 150.1180 &  2.2919 &  7.82 &  1.65 &  4.68 &  0.76 &  0.23 & 150.11828 &   2.29213 &  1.42 &  3.59 &  0.99\\
S2CLSc37.2 & 150.1180 &  2.2919 &  7.82 &  1.65 &  4.68 &  0.76 &  0.23 & 150.11782 &   2.29234 &  1.82 &  2.37 &  1.35\\
S2CLSc38 & 150.1308 &  2.3143 &  6.16 &  1.33 &  4.62 &  1.15 &  0.20 & 150.13074 &   2.31408 &  0.76 &  4.15 &  2.01\\
S2CLSc39 & 150.1464 &  2.3374 &  6.27 &  1.34 &  4.61 &  0.73 &  0.21 &{\it  150.14722 }& {\it   2.33719} &  {\it 2.89} & {\it -0.11} &  {\it 0.72}\\
S2CLSc40 & 150.1352 &  2.3702 &  7.01 &  1.52 &  4.51 &  1.06 &  0.24 & 150.13611 &   2.37052 &  3.37 &  1.78 &  0.81\\
S2CLSc41 & 150.1616 &  2.2684 &  9.97 &  2.17 &  4.49 &  0.99 &  0.33 & {\it 150.16269 }&   {\it 2.26849 }&  {\it 3.93 }&  {\it 0.45 }&  {\it 1.31}\\
S2CLSc42 & 150.1834 &  2.3899 &  8.56 &  1.88 &  4.45 &  0.90 &  0.35 & 150.18368 &   2.39047 &  2.23 &  3.59 &  1.80\\
S2CLSc43 & 150.1620 &  2.3408 &  6.84 &  1.56 &  4.36 &  1.03 &  0.26 & 150.16186 &   2.34092 &  0.64 &  1.20 &  0.66\\
S2CLSc44 & 150.1321 &  2.4024 &  8.29 &  1.89 &  4.33 &  1.63 &  0.29 & 150.13188 &   2.40320 &  2.87 &  3.04 &  3.03\\
S2CLSc45 & 150.0601 &  2.3022 &  9.19 &  2.11 &  4.33 &  0.05 &  0.27 & {\it 150.05904 }&  {\it  2.30241} &  {\it 3.69} &{\it  -0.11 }&  {\it 0.81}\\
S2CLSc46 & 150.0672 &  2.3719 &  9.30 &  2.11 &  4.32 &  0.06 &  0.29 & 150.06710 &   2.37243 &  2.02 &  3.17 &  1.20\\
S2CLSc47 & 150.1770 &  2.3713 &  7.90 &  1.84 &  4.27 &  0.54 &  0.34 & 150.17703 &   2.37144 &  0.63 &  2.61 &  0.68\\
S2CLSc48 & 150.0846 &  2.2545 & 10.24 &  2.37 &  4.25 &  1.20 &  0.32 & {\it 150.08524} &  {\it  2.25521 }&  {\it 3.43 }&  {\it 0.94} & {\it  1.01}\\
S2CLSc49 & 150.1296 &  2.2422 & 10.94 &  2.58 &  4.22 &  0.46 &  0.37 & 150.12956 &   2.24119 &  3.65 &  1.43 &  1.75\\
S2CLSc50 & 150.1434 &  2.4160 &  9.56 &  2.25 &  4.21 &  0.99 &  0.36 & 150.14266 &   2.41605 &  2.59 &  1.88 &  1.33\\
S2CLSc51 & 150.1874 &  2.3799 &  7.87 &  1.84 &  4.21 &  0.12 &  0.34 & 150.18719 &   2.38015 &  1.22 &  3.82 &  2.34\\
S2CLSc52 & 150.1677 &  2.2983 &  7.21 &  1.69 &  4.17 &  1.58 &  0.31 & 150.16771 &   2.29876 &  1.54 &  4.04 &  1.31\\
S2CLSc53 & 150.1660 &  2.3076 &  6.72 &  1.58 &  4.16 &  0.45 &  0.28 & 150.16613 &   2.30747 &  0.63 &  3.48 &  2.85\\
S2CLSc54 & 150.1941 &  2.3113 &  7.76 &  1.86 &  4.14 &  0.16 &  0.33 & 150.19479 &   2.31181 &  3.07 &  1.80 &  0.79\\
S2CLSc55 & 150.1299 &  2.2531 &  9.07 &  2.18 &  4.12 &  0.97 &  0.30 & 150.13001 &   2.25269 &  1.57 &  3.90 &  1.78\\
S2CLSc56 & 150.0683 &  2.2759 &  8.29 &  1.97 &  4.10 &  1.72 &  0.29 & 150.06811 &   2.27569 &  1.08 &  1.96 &    -- \\
S2CLSc57 & 150.1521 &  2.3344 &  5.55 &  1.39 &  4.00 &  0.04 &  0.23 & 150.15182 &   2.33461 &  1.21 &  3.60 &  1.20\\
S2CLSc57.2 & 150.1521 &  2.3344 &  5.55 &  1.39 &  4.00 &  0.04 &  0.23 & 150.15219 &   2.33458 &  0.60 &  2.93 &    -- \\
S2CLSc57.3 & 150.1521 &  2.3344 &  5.55 &  1.39 &  4.00 &  0.04 &  0.23 & 150.15167 &   2.33380 &  2.79 &  1.38 &  1.25\\
S2CLSc57.4 & 150.1521 &  2.3344 &  5.55 &  1.39 &  4.00 &  0.04 &  0.23 & 150.15202 &   2.33361 &  2.98 &  1.19 &  0.75\\
\hline
\end{longtable}

\twocolumn
\begin{table*}
\caption{450$\,\mu$m sources in UDS CANDELS region. Sources with only unreliable identifications (i.e. $\log_{10}LR<1.1$) are shown in italics. Where multiple good identifications for sources exist these are listed with the suffix .2, .3, etc.}
\label{tab:s2uds}
\begin{tabular}{lllllllllllll}
\hline
 ID & RA$_{450}$ & Dec$_{450}$ & $S_{450}$ & $\Delta S_{450}$ & SNR$_{450}$ &  $S_{850}$ & $\Delta S_{850}$ & RA$_{\rm opt}$ & Dec$_{\rm opt}$ & Sep. & $\log_{10}LR$ & $z_{\rm phot}$  \\
  \hline
  & deg. & deg. & mJy & mJy & & mJy & mJy & deg. & deg. & arcsec & & \\
\hline\hline
S2CLSu01 &  34.3639 &  $-$5.1991 & 21.63 &  2.83 &  7.51 &   5.46 &  0.38 &  34.3632 & $-$5.19937 &  2.80 &  3.20 &  2.22\\
S2CLSu01.2 &  34.3639 &  $-$5.1991 & 21.63 &  2.83 &  7.51 &   5.46 &  0.38 &  34.3645 & $-$5.19901 &  2.11 &  2.49 &  2.17\\
S2CLSu02 &  34.3341 &  $-$5.2182 & 27.60 &  4.16 &  6.47 &   5.34 &  0.51 &  34.3334 & $-$5.21825 &  2.43 &  3.37 &  2.15\\
S2CLSu03 &  34.3859 &  $-$5.1987 & 17.31 &  2.81 &  6.04 &   4.27 &  0.38 &  34.3857 & $-$5.19898 &  1.06 &  4.46 &  2.15\\
S2CLSu03.2 &  34.3859 &  $-$5.1987 & 17.31 &  2.81 &  6.04 &   4.27 &  0.38 &  34.3854 & $-$5.19869 &  1.82 &  2.11 &  2.15\\
S2CLSu03.3 &  34.3859 &  $-$5.1987 & 17.31 &  2.81 &  6.04 &   4.27 &  0.38 &  34.3863 & $-$5.19851 &  1.91 &  1.82 &   --\\
S2CLSu03.4 &  34.3859 &  $-$5.1987 & 17.31 &  2.81 &  6.04 &   4.27 &  0.38 &  34.3866 & $-$5.19887 &  2.80 &  1.66 &  2.12\\
S2CLSu04 &  34.4077 &  $-$5.2630 & 27.36 &  4.71 &  5.79 &   3.00 &  0.64 & {\it 34.4068} & {\it$-$5.26278} &  {\it3.59} &{\it $-$0.03} & {\it 2.58}\\
S2CLSu05 &  34.3721 &  $-$5.1977 & 15.86 &  2.80 &  5.51 &   1.95 &  0.37 &  34.3721 & $-$5.19791 &  0.89 &  3.74 &  2.58\\
S2CLSu06 &  34.3788 &  $-$5.1824 & 12.87 &  2.90 &  4.43 &   0.87 &  0.38 &  {\it34.3795}& {\it$-$5.18232} &  {\it2.78} &{\it $-$2.18} & {\it 1.07}\\
S2CLSu07 &  34.3073 &  $-$5.1612 & 24.36 &  5.70 &  4.22 &   0.41 &  0.66 &  34.3067 & $-$5.16106 &  1.97 &  3.75 &  1.62\\
S2CLSu07.2 &  34.3073 &  $-$5.1612 & 24.36 &  5.70 &  4.22 &   0.41 &  0.66 &  34.3071 & $-$5.16146 &  1.30 &  3.15 &  1.40\\
S2CLSu07.3 &  34.3073 &  $-$5.1612 & 24.36 &  5.70 &  4.22 &   0.41 &  0.66 &  34.3074 & $-$5.16122 &  0.61 &  2.83 &  1.45\\
S2CLSu08 &  34.4326 &  $-$5.2390 & 18.73 &  4.40 &  4.14 &   0.65 &  0.69 &  34.4322 & $-$5.23956 &  2.37 &  2.01 &   --\\
S2CLSu09 &  34.3244 &  $-$5.2171 & 19.59 &  4.64 &  4.13 &  $-$0.43 &  0.53 & {\it 34.3248} & {\it$-$5.21710} &{\it  1.34} & {\it$-$0.86} &   --\\
S2CLSu10 &  34.4601 &  $-$5.2013 & 21.62 &  5.24 &  4.11 &   2.88 &  0.80 &  34.4612 & $-$5.20134 &  3.92 &  1.95 &  2.07\\
S2CLSu11 &  34.3031 &  $-$5.2192 & 22.93 &  5.43 &  4.02 &   2.27 &  0.58 &  34.3022 & $-$5.21920 &  3.06 &  2.23 &  2.20\\
S2CLSu12 &  34.3384 &  $-$5.2532 & 17.83 &  4.37 &  4.01 &   2.89 &  0.55 &  34.3382 & $-$5.25408 &  3.26 &  1.62 &  0.82\\
\hline
\end{tabular}
\end{table*}

\begin{table*}
\caption{Multi-wavelength photometry and derived quantities for the SCUBA-2 450$\,\mu$m sources with robust counterparts and photometric redshifts.}
\label{tab:s2details}
\begin{tabular}{lrrrrrrrrrlrrr}
\hline
ID & $S_{250}$ & $\Delta S_{250}$ &  $S_{350}$ & $\Delta S_{350}$ &  $S_{450}$ & $\Delta S_{450}$ &  $S_{850}$ & $\Delta S_{850}$ & $L_{\rm IR}$ & $T_{\rm K}$ & $\beta_{\rm D}$  & $M_{\ast}$ & $R_{e}$  \\
\hline
 & mJy & mJy & mJy & mJy & mJy & mJy & mJy & mJy & L$_{\odot}$ & K & & M$_{\odot}$ & kpc\\
\hline\hline
c01 &  42.3 &   3.4 &  38.5 &   4.9 &  24.5 &   1.8 &   3.2 &   0.6 & $ 12.8\pm  0.2$ & $ 40.6^{+  3.7}_{-  3.2}$ & $ 2.51^{+  0.2}_{-  0.2}$ &  11.3 &  0.47\\
c02 &  26.5 &   3.4 &  35.8 &   5.1 &  18.9 &   1.4 &   5.2 &   0.5 & $ 12.9\pm  0.3$ & $ 54.4^{+ 10.1}_{-  6.9}$ & $ 1.38^{+  0.3}_{-  0.3}$ &  10.5 &  0.47\\
c03 &  22.5 &   3.4 &  22.0 &   4.8 &  15.2 &   1.4 &   4.7 &   0.6 & $ 12.8\pm  0.2$ & $ 56.0^{+  6.0}_{-  6.5}$ & $ 0.94^{+  0.2}_{-  0.2}$ &  10.2 &  0.28\\
c04 &  19.7 &   3.4 &  22.6 &   4.8 &  14.7 &   1.4 &   4.2 &   0.5 & $ 12.8\pm  0.3$ & $ 53.5^{+ 13.2}_{-  8.6}$ & $ 1.20^{+  0.4}_{-  0.4}$ &  10.7 &  0.04\\
c05 &  20.2 &   3.4 &  29.4 &   4.9 &  16.7 &   1.7 &   5.6 &   0.6 & $ 12.8\pm  0.1$ & $ 53.4^{+  5.7}_{-  5.5}$ & $ 0.96^{+  0.3}_{-  0.2}$ &  10.8 &  0.45\\
c06 &  17.1 &   3.4 &  26.8 &   4.9 &  20.9 &   2.3 &  10.4 &   0.6 & $ 12.0\pm  0.1$ & $ 25.9^{+  2.8}_{-  3.7}$ & $ 0.84^{+  0.4}_{-  0.2}$ &  11.2 &  0.50\\
c07 &  14.6 &   3.4 &  26.5 &   5.0 &  14.7 &   1.6 &   6.9 &   0.6 & $ 10.7\pm  0.2$ & $ 16.8^{+  2.0}_{-  2.6}$ & $ 0.78^{+  0.4}_{-  0.2}$ &  10.7 &  0.17\\
c08 &  39.0 &   3.4 &  36.4 &   4.8 &  18.1 &   2.2 &   3.5 &   0.6 & $ 12.9\pm  0.2$ & $ 49.9^{+  4.2}_{-  4.2}$ & $ 1.92^{+  0.2}_{-  0.2}$ &  10.2 &  0.55\\
c09 &  26.0 &   3.4 &  25.0 &   4.8 &  18.2 &   2.3 &   3.4 &   0.6 & $ 11.9\pm  0.3$ & $ 22.7^{+  3.9}_{-  2.8}$ & $ 2.04^{+  0.4}_{-  0.4}$ &  10.6 & -0.05\\
c10 &  11.1 &   3.4 &  17.0 &   4.9 &  12.0 &   1.6 &   2.8 &   0.6 & $ 12.4\pm  0.4$ & $ 36.1^{+ 10.4}_{-  7.0}$ & $ 1.95^{+  0.6}_{-  0.6}$ &  11.2 &  0.60\\
c11 &  19.0 &   3.4 &  15.8 &   5.2 &  12.0 &   1.6 &   2.2 &   0.6 & $ 12.0\pm  0.4$ & $ 34.0^{+ 21.8}_{-  8.1}$ & $ 1.74^{+  0.6}_{-  0.7}$ &  12.5 &  0.55\\
c12 &  12.7 &   3.4 &  15.0 &   4.9 &  10.1 &   1.4 &   2.9 &   0.5 & $ 12.6\pm  0.4$ & $ 51.0^{+ 14.9}_{- 12.1}$ & $ 1.22^{+  0.6}_{-  0.5}$ &  11.2 &  0.42\\
c13 &   9.8 &   3.4 &  25.6 &   6.2 &  13.4 &   1.9 &   3.2 &   0.6 & $ 12.2\pm  0.4$ & $ 28.4^{+  6.9}_{-  4.3}$ & $ 2.20^{+  0.5}_{-  0.6}$ &  11.2 &  0.59\\
c15 &  13.0 &   3.4 &  13.4 &   4.9 &  10.6 &   1.6 &   2.2 &   0.6 & $ 12.2\pm  0.4$ & $ 36.3^{+ 18.2}_{-  9.0}$ & $ 1.64^{+  0.7}_{-  0.7}$ &  11.0 &  0.53\\
c16 &  10.3 &   3.4 &  15.0 &   4.9 &   8.6 &   1.4 &   2.0 &   0.5 & $ 12.1\pm  0.5$ & $ 39.3^{+ 20.8}_{- 10.7}$ & $ 1.52^{+  0.7}_{-  0.7}$ &  10.4 &  0.53\\
c17 &  17.5 &   3.4 &  13.3 &   4.9 &   8.3 &   1.3 &   0.2 &   0.5 & $ 11.9\pm  0.1$ & $ 46.4^{+  2.2}_{-  2.4}$ &  --  &  11.5 &  0.67\\
c18 &  10.3 &   3.4 &   7.5 &   4.8 &   7.5 &   1.2 &   1.8 &   0.5 & $ 12.3\pm  0.5$ & $ 50.6^{+ 24.1}_{- 15.9}$ & $ 1.21^{+  0.8}_{-  0.5}$ &  11.2 &  0.37\\
c19 &  18.2 &   3.4 &  14.7 &   4.9 &  10.7 &   1.8 &   1.3 &   0.6 & $ 12.5\pm  0.1$ & $ 53.5^{+  5.7}_{-  5.5}$ &  --  &  11.1 &  0.39\\
c20 &  26.6 &   3.4 &  16.2 &   5.0 &   9.9 &   1.8 &   1.1 &   0.6 & $ 12.1\pm  0.1$ & $ 37.5^{+  1.3}_{-  1.6}$ &  --  &  11.4 &  0.32\\
c21 &   5.3 &   3.4 &   5.8 &   4.9 &   7.1 &   1.3 &   1.2 &   0.5 & $ 11.4\pm  0.3$ & $ 40.5^{+ 14.6}_{- 10.3}$ &  --  &   9.4 &  0.62\\
c22 &  10.1 &   3.4 &   2.7 &   5.1 &  11.6 &   2.1 &   2.9 &   0.6 & $ 11.7\pm  0.1$ & $ 27.2^{+  2.7}_{-  2.7}$ &  --  &  10.6 &  0.40\\
c23 &  10.2 &   3.4 &   6.7 &   4.8 &  11.4 &   2.1 &   1.8 &   0.6 & $ 10.5\pm  0.3$ & $ 32.3^{+  8.7}_{-  6.2}$ &  --  &   8.5 & -0.06\\
c25 &   2.4 &   3.4 &   0.0 &  12.2 &   8.4 &   1.7 &   1.9 &   0.6 & $ 10.4\pm  0.2$ & $ 18.8^{+  3.3}_{-  3.1}$ &  --  &   9.4 &  0.50\\
c26 &   8.2 &   3.4 &  11.2 &   5.4 &   9.2 &   1.8 &   2.8 &   0.6 & $ 11.9\pm  0.1$ & $ 30.0^{+  3.0}_{-  3.3}$ &  --  &  10.7 &  0.25\\
c28 &   0.0 &   3.4 &   0.0 &   4.8 &   9.7 &   1.9 &  -0.2 &   0.6 & $ 11.6\pm   0.3 $ & $ 42.0^{+ 35.1}_{- 31.0}$ &  --  &   9.6 &  0.55\\
c29 &   6.6 &   3.4 &  12.2 &   5.1 &   8.7 &   1.7 &   1.3 &   0.6 & $ 12.1\pm  0.1$ & $ 40.3^{+  5.2}_{-  4.8}$ &  --  &  11.1 &  0.44\\
c30 &  11.3 &   3.4 &  14.0 &   4.9 &   8.3 &   1.7 &   0.0 &   0.6 & $ 12.0\pm  0.2$ & $ 78.1^{+ 15.2}_{- 27.0}$ &  --  &   9.9 &  0.32\\
c31 &   4.4 &   3.4 &   9.8 &   5.1 &   5.9 &   1.2 &   1.6 &   0.5 & $ 12.1\pm  0.1$ & $ 36.2^{+  4.9}_{-  5.1}$ &  --  &  10.9 &  0.45\\
c34 &   6.8 &   3.4 &   4.9 &   4.8 &   8.9 &   1.8 &   2.2 &   0.6 & $  8.8\pm  0.2$ & $ 18.0^{+  3.6}_{-  3.1}$ &  --  &   8.4 & -0.93\\
c35 &  13.3 &   3.4 &   8.0 &   6.9 &   8.2 &   1.7 &   3.0 &   0.6 & $ 12.4\pm  0.4$ & $ 51.0^{+ 11.8}_{- 12.0}$ & $ 0.84^{+  0.5}_{-  0.3}$ &  11.1 &  0.47\\
c37 &   9.8 &   3.4 &   7.8 &   4.9 &   7.8 &   1.7 &   0.8 &   0.6 & $ 11.4\pm  0.2$ & $ 25.6^{+  2.8}_{-  2.8}$ &  --  &  11.8 &  0.52\\
c38 &   9.7 &   3.4 &   5.2 &   5.2 &   6.2 &   1.3 &   1.1 &   0.5 & $ 12.3\pm  0.2$ & $ 50.7^{+  8.5}_{-  8.4}$ &  --  &  10.8 &  0.34\\
c39 &  22.3 &   3.4 &  11.4 &   5.2 &   6.3 &   1.3 &   0.7 &   0.5 & $ 11.7\pm  0.1$ & $ 40.7^{+  1.3}_{-  1.5}$ &  --  &  10.9 &  0.32\\
c40 &  22.3 &   3.4 &  18.5 &   5.1 &   7.0 &   1.5 &   1.1 &   0.6 & $ 11.9\pm  0.1$ & $ 31.5^{+  1.0}_{-  1.0}$ &  --  &  12.0 &  0.35\\
c42 &  10.4 &   3.4 &  12.0 &   4.9 &   8.6 &   1.9 &   0.9 &   0.6 & $ 12.4\pm  0.2$ & $ 61.8^{+ 13.7}_{- 10.8}$ &  --  &  10.7 &  0.61\\
c43 &  12.8 &   3.4 &   5.3 &   4.8 &   6.8 &   1.6 &   1.0 &   0.6 & $ 11.0\pm  0.1$ & $ 17.6^{+  1.3}_{-  1.4}$ &  --  &  12.8 &  0.82\\
c46 &   7.8 &   3.4 &   6.1 &   5.0 &   9.3 &   2.1 &   0.1 &   0.6 & $ 11.9\pm  0.4$ & $ 71.3^{+ 19.6}_{- 29.7}$ &  --  &  11.2 &  0.32\\
c47 &   0.0 &   8.8 &   3.6 &   4.8 &   7.9 &   1.8 &   0.5 &   0.6 & $ 11.5\pm   0.3 $ & $ 66.8^{+ 22.3}_{- 29.9}$ &  --  &   9.2 &  0.60\\
c49 &   0.0 &   4.9 &   0.0 &   9.5 &  10.9 &   2.6 &   0.5 &   0.6 & $ 12.6\pm  0.5$ & $ 75.4^{+ 16.4}_{- 25.1}$ &  --  &  11.4 &  0.54\\
c50 &   0.0 &   7.2 &   0.0 &   8.6 &   9.6 &   2.3 &   1.0 &   0.6 & $ 12.3\pm  0.5$ & $ 75.3^{+ 16.6}_{- 24.8}$ &  --  &  10.0 &  0.21\\
c51 &  10.6 &   3.4 &  12.6 &   4.9 &   7.9 &   1.8 &   0.1 &   0.6 & $ 12.4\pm  0.2$ & $ 86.2^{+  9.8}_{- 18.1}$ &  --  &  11.9 &  0.32\\
c52 &   7.7 &   3.4 &  11.0 &   5.0 &   7.2 &   1.7 &   1.6 &   0.6 & $ 11.6\pm  0.2$ & $ 30.1^{+  4.1}_{-  3.8}$ &  --  &  10.6 &  0.42\\
c53 &   3.0 &   3.4 &   4.1 &   4.9 &   6.7 &   1.6 &   0.5 &   0.6 & $ 12.3\pm  0.3$ & $ 72.8^{+ 18.0}_{- 23.4}$ &  --  &  10.5 &  0.03\\
c54 &  14.4 &   3.4 &   6.1 &   4.9 &   7.8 &   1.9 &   0.2 &   0.6 & $ 11.4\pm  0.1$ & $ 51.1^{+  3.6}_{-  4.2}$ &  --  &  10.8 &  0.57\\
c55 &  13.5 &   3.4 &  10.9 &   4.9 &   9.1 &   2.2 &   1.0 &   0.6 & $ 12.6\pm  0.3$ & $ 77.4^{+ 13.2}_{- 16.1}$ &  --  &  10.4 &  0.29\\
c57 &  10.0 &   3.4 &   7.5 &   5.3 &   5.5 &   1.4 &   0.0 &   0.5 & $ 12.1\pm  0.5$ & $ 76.4^{+ 16.3}_{- 25.5}$ &  --  &  12.0 &  0.53\\
u01 &  36.7 &   3.6 &  38.7 &   4.0 &  21.6 &   2.8 &   5.5 &   0.6 & $ 12.9\pm  0.3$ & $ 47.5^{+ 10.0}_{-  6.6}$ & $ 1.58^{+  0.3}_{-  0.4}$ &  11.4 &  0.62\\
u02 &  38.9 &   3.6 &  43.7 &   4.1 &  27.6 &   4.2 &   5.3 &   0.7 & $ 12.9\pm  0.3$ & $ 35.9^{+  4.8}_{-  3.3}$ & $ 2.31^{+  0.3}_{-  0.3}$ &  11.4 &  0.64\\
u03 &  37.9 &   3.6 &  35.4 &   4.1 &  17.3 &   2.8 &   4.3 &   0.6 & $ 12.9\pm  0.5$ & $ 56.3^{+ 19.2}_{- 10.5}$ & $ 1.45^{+  0.4}_{-  0.4}$ &  11.0 &  0.59\\
u05 &  13.7 &   3.6 &  15.0 &   4.1 &  15.9 &   2.8 &   3.9 &   0.6 & $ 12.5\pm  0.4$ & $ 40.6^{+ 12.5}_{-  8.3}$ & $ 1.70^{+  0.7}_{-  0.7}$ &  10.5 &  0.16\\
u07 &  37.1 &   3.6 &  22.2 &   4.1 &  24.4 &   5.7 &   0.4 &   0.8 & $ 13.0\pm  0.5$ & $ 73.8^{+ 15.9}_{- 11.2}$ & $ 2.80^{+  0.1}_{-  0.2}$ &  11.4 &  0.33\\
u10 &  43.8 &   3.6 &  35.2 &   6.0 &  21.6 &   5.2 &   2.9 &   0.9 & $ 12.9\pm  0.5$ & $ 47.6^{+ 21.3}_{-  8.5}$ & $ 2.20^{+  0.4}_{-  0.6}$ &  11.0 &  0.49\\
u11 &  13.6 &   3.6 &  12.5 &   5.2 &  22.9 &   5.4 &   2.3 &   0.8 & $ 12.5\pm  0.4$ & $ 39.8^{+ 21.1}_{-  8.9}$ & $ 1.89^{+  0.7}_{-  0.9}$ &  11.2 &  0.38\\
u12 &   5.6 &   3.6 &   7.8 &   4.9 &  17.8 &   4.4 &   2.9 &   0.7 & $ 11.1\pm  0.2$ & $ 18.9^{+  2.4}_{-  2.5}$ &  --  &  10.4 &  0.14\\

\hline
\end{tabular}
\end{table*}

\section{MCMC fitting technique}\label{app:mcmc}
The modified blackbody parameters are inferred from the observed SEDs using a Markov Chain Monte Carlo (MCMC) methodology. Specifically, the posterior probability density is mapped using a Metropolis-hastings algorithm with Gibbs sampling. Parameters are constrained to take realistic values with a top-hat prior. The parameter constraints used in this work are given in Table \ref{tab:mcmclims}.

\begin{table}
\caption{Limits on the parameters in the SED fitting}
\label{tab:mcmclims}
\begin{tabular}{lrr}
\hline
Parameter & Min & Max \\
\hline\hline
$\log_{10}L_{\rm IR}$ $(L_{\odot})$ & 9 & 14\\
$T_{\rm D} (K)$ & 5 & 100\\
$\beta_{\rm D}$ & 0.5 & 3.5\\
$\nu_0$ & 100 & 100\\
\hline
\end{tabular}
\end{table}

The likelihood is estimated via use of the $\chi^2$ statistic. Ten parallel MCMC chains are run, with convergence checks performed every 400 iterations, after an initial burn-in of 2000 iterations. Convergence is assessed using the method of Gelman \& Rubin (1992). After convergence is achieved a further 2000 iterations are performed, with inference performed on the last 2200 iterations from each chain.
\section{Stacking for confusion-limited submm imaging}\label{sec:stacking}
As multi-wavelength coverage of extra-galactic survey fields has become common-place, the practice of stacking images based on a selection at another, potentially deeper, wavelength has become popular. However, a major difficulty in this approach is the mis-matched resolution between data sets, e.g. the optical and submm datasets used in this work have angular resolutions that are different by a factor of roughly $100$. This means that the ``raw'' stacked signal (i.e. just the pixels combined together) may not only contain signal from the sources in the input list, but also those that are correlated spatially with them. It is important to note that the corruption of the stacking signal is solely due to these correlations; if galaxies were purely Poisson distributed then simply measuring the co-variance between the pixels coincident with sources would give an unbiased estimate of their mean flux density (e.g.\ Marsden et al.\ 2009).

Recently, a number of authors have presented methods to deal with these correlations. B{\'e}thermin et al.\ (2012a) use a traditional stacking method, but correct for the clustering signal via the use of simulations. Meanwhile Kurczynski \& Gawiser (2010; KG10) offer an approach which uses linear algebra to account for the correlations between sources in the input list. Other methodologies exist but mainly fall into these two categories. Both of these methods have significant drawbacks. In principle the B{\'e}thermin et al.\ (2012a) method can completely correct for the clustered signal, but it requires perfect knowledge of the clustering properties of the galaxies in the stacking list, something which is difficult to obtain given that we know clustering is a function of intrinsic galaxy properties (e.g. star formation rate, stellar mass, morphology, etc). The KG10 approach again offers the ability to completely correct the estimates, but this time requires the stacking input list to be complete; only correlations with galaxies in the input list can be taken into account. Thus the two approaches suit different applications; if the source list to be stacked is incomplete but homogeneous then the first approach is better, if it is quite complete but potentially heterogenous then the second approach should be used.

As an aside, a small number of papers have suggested that the best method to remove correlations is to fit for bright sources and then stack ``source-free'' maps (e.g. Chary, Cooray \& Sullivan 2008). However, this approach requires perfect knowledge of the source photometry, otherwise it is likely to introduce artifacts that can become significant if correlations between the bright sources and faint stacking list exist. 

Given the depth of the CANDELS imaging available in the UDS and COSMOS fields we make use of a KG10-like approach to stack the submm imaging, although with some mathematical tweaks. 

To begin, we assume that the PRF is circularly symmetric and so can be described in one dimension as a function of the radial distance from the source position. This assumption is for simplicity and to reduce the computational expense of the stacking calculation; all the following formalism can be easily expanded to consider the full 2D stacked image. For a 1D stack the flux density for a single source $x$ at a distance $r$ is given by,

\[
  s(r)=1/N_r\sum d(x+r),
\]

\noindent where $d$ is the image, and the sum runs over the $N_r$ pixels that lie at a separation $r$ from the source position. It is clear that the simplest 1D stack for a list of source positions ${\bf x}$ is then,
\[ 
\bar{s(r)}=\frac{1}{N_rN_x}\sum_r\sum_x d(x+r).
\]

If image pixels in $d$ appear more than once in the summation then this estimation for the stacked 1D source profile will be biased, giving erroneously high values of $\bar{s(r)}$ and broader source profiles. To correct this we need to include information about the correlations between the pixels to be stacked. We can re-write the above equation in matrix notation as,

\begin{equation}
{\bf \bar{s}}={\bf Ad},
\label{eqn:naivestack}
\end{equation}

\noindent where ${\bf A}$ is a $N_d\times N_r$ matrix which contains the value $1/(N_xN_r)$ at the pixel coordinate that corresponds to a source $x$ and separation $r$. For a source list without correlations, the columns of ${\bf A}$ would contain at most one non-zero value. This means that for an uncorrelated source list ${\bf A^TA}={\bf I}$, where ${\bf I}$ is the identity matrix. If correlations between sources in the list exist then ${\bf A^TA}$ will be a symmetric matrix that describes these correlations. Thus it can be shown that an unbiased estimate of the mean 1D profile is given by 

\begin{equation}
{\bf A^TA\bar{s}}={\bf Ad}.
\end{equation}

Letting ${\bf A^TA}={\bf P}$ and ${\bf Ad}={\bf b}$, this is now a linear equation of the form ${\bf b=Ps}$, which can be solved efficiently via conjugate gradient methods (as the matrix ${\bf P}$ is non-negative and symmetric). In this formalism ${\bf b}$ can be interpreted as the ``naive stack'' i.e. the stack we would get by simply adding up the pixel flux from the map for each source, while ${\bf P}$ is the Fisher information matrix. Measurement errors in the image can be included naturally by re-writing ${\bf P}$ and ${\bf b}$ as,
\[
\begin{array}{l}
{\bf P=A^TN^{-1}A},\\
{\bf b=A^TN^{-1}d}.\\
\end{array}
\]

For stacking to give interpretable results the source lists that go into the stack must be reasonably homogenous. This creates a tension with our desire to stack as complete a sample as possible to account for all the correlations that may be present. Fortunately the stacking methodology can be expanded so that any number of independent source lists can be stacked simultaneously to account for the correlations both within lists and between them.

Assuming now we have a set of source positions to be stacked $\bar{{\bf s}^1}, \bar{{\bf s}^2}...\bar{{\bf s}^N}$, we can create matrices ${\bf A}^1, {\bf A}^2....{\bf A}^N$ that map the $d$ into these $N_t$ stacks. We can then re-write Eq.~\ref{eqn:naivestack} as
\[
\begin{array}{cccc}
 & & [{\bf A}^1  &\\
{[ \bar{\bf{s}}^1, \bar{ \bf{s}}^2 , ..., \bar{\bf{s}^N}]} & = & {\bf A}^2 & {\bf d} \\
 &  & ... &  \\
& &  {\bf A}^N] & \\
\end{array}
\]
Next we calculate ${\bf P}$ and ${\bf b}$ as before, although now ${\bf P}$ is a $N_rN_t\times N_rN_t$ matrix and the solution vector $\bar{\bf{s}}$ is length $N_rN_t$ and contains the $N_r$ 1D profiles for the $N_t$ stacks.


\end{document}